\documentclass[10pt,a4paper]{article}
\pdfoutput=1
\usepackage{wrapfig}

\usepackage[margin=1in]{geometry}
\usepackage{amsmath,amssymb,bm}
\usepackage{hyperref}
\usepackage{xcolor} 
\usepackage{graphicx,rotating,hyperref,slashed,amsmath,xcolor,amssymb,amsfonts,colortbl,cite, subfigure,float}
\usepackage{setspace}
\numberwithin{equation}{section}
\usepackage{mciteplus}

\hypersetup{colorlinks,bookmarksopen,bookmarksnumbered,
linkcolor=blue,pdfstartview=FitH,urlcolor=blue,citecolor=blue}

\hypersetup{%
    ,urlcolor=black
    ,citecolor=black
    ,linkcolor=black
    }
\usepackage{mciteplus} 

\usepackage{multicol}
\usepackage{color}
\definecolor{rosso}{cmyk}{0,1,1,0.4}
\definecolor{rossos}{cmyk}{0,1,1,0.55}
\definecolor{rossoc}{cmyk}{0,1,1,0.2}
\definecolor{blu}{cmyk}{1,1,0,0.3}
\definecolor{blus}{cmyk}{1,1,0,0.6}
\definecolor{bluc}{cmyk}{1,1,0,0.1}
\definecolor{verde}{cmyk}{0.92,0,0.59,0.25}
\definecolor{verdec}{cmyk}{0.92,0,0.59,0.15}
\definecolor{verdes}{cmyk}{0.92,0,0.59,0.4}

\def\circa#1{\,\raise.3ex\hbox{$#1$\kern-.75em\lower1ex\hbox{$\sim$}}\,}

\newcommand{\be}{\begin{equation}}
\newcommand{\ee}{\end{equation}}
\newfam\rsfsfam
\def\mathscr#1{{\fam\rsfsfam\relax#1}}

\def\circa#1{\,\raise.3ex\hbox{$#1$\kern-.75em\lower1ex\hbox{$\sim$}}\,}
\makeatletter

\hypersetup{%
    ,urlcolor=black
    ,citecolor=black
    ,linkcolor=black
    }
\usepackage{mciteplus}

\AtBeginDocument{
  \hypersetup{
    urlcolor=blue,
    citecolor=blue,
    linkcolor=black,
  }%
}

\newcommand{\dd}{\mathrm{d}}

\begin{document}

\setcounter{page}{1} \baselineskip=15.5pt \thispagestyle{empty}

\vspace{0.8cm}
\begin{center}

{\fontsize{19}{28}\selectfont  \sffamily \bfseries {
Relaxation without Ringdown for a  \\ \vskip0.25cm  Compact Object in Modified Gravity
}}

\end{center}

\vspace{0.2cm}

\begin{center}
{\fontsize{13}{30}\selectfont  Gianmassimo Tasinato$^{1,2}$ } 
\end{center}

\begin{center}

\vskip 8pt
\textsl{$^{1}$Physics Department, Swansea University, SA28PP, United Kingdom
  }
\\
\textsl{$^{2}$ Dipartimento di Fisica e Astronomia, Universit\`a di Bologna,  Italy
 }\\
\textsl{\texttt{email}: g.tasinato2208 at gmail.com }\\
\vskip 7pt

\end{center}

\smallskip
\begin{abstract}
\noindent
Compact objects with black-hole-like exteriors may hide new strong-field physics in their interiors, making their dynamical response a sensitive probe of gravity beyond General Relativity. We present an analytically tractable, gravitationally bound compact object with a genuinely new dynamical signature: under a minimal passive boundary prescription, its exactly controlled odd-parity sector exhibits purely dissipative relaxation poles, rather than the oscillatory modes usually associated with black holes and exotic compact alternatives.
The object we study is a regular, vector-supported compact solution of a vector--tensor theory, matched without any surface layer to an exterior Schwarzschild geometry. Owing to its anisotropic stress, it can violate the Buchdahl bound and be continuously connected to the black-hole compactness limit. Its unusual response follows from a hidden chiral symmetry, which turns the perturbation problem into one-way transport rather than ordinary wave propagation. The exterior region alone has no conventional quasinormal-mode spectrum; instead, the regular interior and the matching conditions break the symmetry and quantize the fluctuation spectrum.
We analytically compute the retarded Green function and susceptibility, and derive an effective membrane response by integrating out the object's interior. In the black-hole limit, the relaxation times diverge, the poles collapse toward zero frequency, and finite-frequency exterior perturbations decouple from the interior. Black-hole behaviour is therefore approached through the disappearance of relaxation modes, not through the emergence of ringdown.
\end{abstract}

\setstretch{1.1}


\section{Introduction}

Gravitational-wave observations have turned the dynamics of compact
objects into a precision probe of strong-field gravity.  In General
Relativity, the late-time response of a perturbed black hole is governed
by quasinormal modes: a discrete set of damped oscillations selected by
ingoing boundary conditions at the horizon and outgoing boundary
conditions at infinity
\cite{Nollert:1999ji,Kokkotas:1999bd,Berti:2009kk,Konoplya:2011qq}.
This ringdown paradigm is one of the basis of black-hole
spectroscopy.  It relies on a specific  property of
the perturbation problem: the existence of a two-way wave equation, with
independent ingoing and outgoing branches related by global boundary
conditions.  It is therefore natural to ask how robust this picture is
for compact objects beyond General Relativity, especially for objects
whose exterior geometry can be indistinguishable from Schwarzschild.

\smallskip

This question is closely connected with the broader program of using
compact-object spectroscopy and tidal response to distinguish black holes
from exotic or modified-gravity alternatives.  The phenomenology of
exotic compact objects has been extensively studied, including echoes,
modified ringdown spectra, tidal response, and dissipative effects
\cite{Cardoso:2016rao,Abedi:2016hgu,Cardoso:2016oxy,Cardoso:2017cqb,Conklin:2017lwb,Mark:2017dnq,Wang:2018gin,Wang:2019rcf,Tsang:2019zra,Uchikata:2019frs,Maggio:2019zyv,Maggio:2020jml};
see also the reviews
\cite{Cardoso:2019rvt,Berti:2025hly,Rodriguez:2026iot,Chakraborty:2026qru}.
Most discussions assume, though, that the relevant
linearized dynamics remains wave-like, so that deviations from the
black-hole prediction appear as shifted quasinormal modes, additional
trapped modes, echoes, or modified boundary conditions.  

\medskip

In this work we
exhibit a new possibility, which can arise
in  theories of modified gravity: a compact object with a Schwarzschild
exterior whose analytically controlled odd-parity response is not a
ringdown problem at all.  Our starting point is
 a regular, gravitationally bound compact object in a
vector--tensor theory with a non-minimal coupling between a vector field
and the Einstein tensor.  The solution consists of a vector-supported
interior matched at a finite radius \(R\) to an exactly Schwarzschild
exterior \cite{Tasinato:2022vop}.  The matching surface carries no
localized stress-energy: no thin shell or membrane is introduced in the
fundamental spacetime description.  Owing to the anisotropic stress
sourced by the vector profile, the object can violate the Buchdahl bound
and its compactness can be continuously increased toward the black-hole
value.  This makes the system a controlled analytic laboratory in which
to ask how black-hole-like behaviour is approached in modified
gravity.\footnote{Several works studied aspects of related
vector--tensor black-hole or relativistic-star systems; see e.g.
\cite{Chagoya:2016aar,Chagoya:2017fyl,Heisenberg:2017xda,Minamitsuji:2017aan,Kase:2017egk,Chagoya:2017ojn,Kase:2018voo,Heisenberg:2018acv,Garcia-Saenz:2021uyv,Atkins:2023axs,Aoki:2023bmz,Tomizuka:2025dpy}.}

Under a minimal passive
boundary prescription, the spin-1 odd sector and the homogeneous spin-2
axial sector possess purely damped relaxation poles rather than
oscillatory quasinormal modes.  For each multipole the intrinsic
frequency is of the form
\begin{equation}
\omega_\star=-i\Gamma_{\rm rel},
\qquad
\Gamma_{\rm rel}>0,
\label{eq_intrst}
\end{equation}
so that the response decays monotonically in time,
\(\sim e^{-\Gamma_{\rm rel}t}\), without ringdown oscillations and
without overtones. The quantities  $\Gamma_{\rm rel}$ can be 
computed, and depend on the physics of the interior. The object therefore provides an  analytic
example of relaxation without ringdown: its exterior is Schwarzschild,
but its odd-parity dynamics is qualitatively different from that of a
Schwarzschild black hole, as well as from other exotic compact objects studied in the literature.

The origin of this behaviour is the specific structure of the evolution
equations.  The odd-parity perturbations
are sensitive to the non-trivial vector condensate supporting the
solution.  In the spin-1 vector sector, and in the homogeneous part of
the spin-2 axial sector, the perturbation equations do not reduce to
ordinary Regge--Wheeler wave equations.  Instead, they factorize into
commuting first-order operators.  In variables adapted to the
characteristic flow, the dynamics becomes transport along a single
direction.  This factorization is the local manifestation of a hidden
chiral symmetry, whose bulk realization contains an
\(SL(2,\mathbb{R})\) conformal subgroup.
This chiral structure has an immediate consequence for the spectral
problem.  The exterior Schwarzschild region, taken by itself, does not
define a conventional quasinormal-mode problem in this sector.  There is
no pair of independent counter-propagating branches on which to impose
the usual ingoing/outgoing conditions.  Spatial infinity is an inflow
boundary for the transport problem, rather than an outflow boundary for
radiation.  The relevant spectral problem is therefore intrinsically
global: it is defined only after including the regular interior, the
matching conditions at \(r=R\), and an asymptotic prescription.  These
global conditions break the bulk chiral symmetry and quantize the
fluctuations for the minimal stable
prescription we adopt.\footnote{This situation is reminiscent of
total-transmission modes in Schwarzschild perturbation theory, where a
special algebraic structure causes one wave amplitude to decouple
\cite{Chandrasekhar:1984mgh,Wald:1973wwa,MaassenvandenBrink:2000iwh}.
In the present theory, however, unidirectionality is a generic feature
of the odd sector for all multipoles, rather than an exceptional property
of isolated frequencies.  See also \cite{Capuano:2026tjy} for a recent
setup where purely imaginary frequencies arise for black holes in General Relativity with
constant-rate mass evolution.}

The same physics appears in the retarded Green function and in the
linear response of the compact object.  Because the perturbation
equations are factorized, the Green function can be constructed
analytically.  Its singular part contains precisely the relaxation pole
in Eq.~\eqref{eq_intrst}, while the remaining contribution is a regular
transport term.  After projection on a smooth external probe profile,
the retarded susceptibility takes the Debye form
\begin{equation}
\chi^R(\omega)=
%
\chi^R_{\rm reg}(\omega)
+
\frac{\mathcal A}{\Gamma_{\rm rel}-i\omega}.
\end{equation}
Thus, in the odd-parity sector, the compact object
responds as an overdamped degree of freedom rather than as a resonant
oscillator.  We also analyze the associated static response.  The
zero-frequency branch coefficient is fixed by regularity at the centre
of the object and is therefore not identical to the usual Schwarzschild
black-hole Love number, whose definition uses horizon regularity from
the outset.

A complementary interpretation is obtained by integrating out the
regular interior.  The full smooth-spacetime problem can be rewritten as
an exterior transport problem supplemented by a frequency-dependent
boundary kernel localized at \(r=R\).  This kernel is the exact
Dirichlet-to-Neumann map of the interior and provides a concrete
realization of an effective membrane description
\cite{Damour:1978cg,Thorne:1986iy,Price:1986yy}.  In this
coarse-grained formulation, the boundary kernel is the explicit
symmetry-breaking term that reduces the bulk chiral symmetry to
translations along the characteristic coordinate.  The relaxation scale
is therefore the dynamical remnant of boundary-induced chiral-symmetry
breaking.

We also examine the black-hole compactness limit.  As the parameters
of the regular interior are tuned so that the compactness approaches
\(\mathcal C=1/2\), the relaxation time in Eq.~\eqref{eq_intrst}
diverges.  Equivalently, the dissipative poles collapse toward
\(\omega=0\), and finite-frequency exterior perturbations decouple from
the interior.  The object  becomes black-hole-like not by
reproducing the Schwarzschild quasinormal-mode tower, but by losing its
 finite-frequency relaxation response.  This provides an analytic
example in which a Schwarzschild exterior hides a dynamical response that
is qualitatively different from ordinary  ringdown.

\smallskip

The paper is organized as follows.  In
Section~\ref{sec:background} we review the vector--tensor theory and the
regular compact-object solution.  In
Section~\ref{sec:perturbations} we derive the odd-parity perturbation
equations, exhibit their factorized form, and identify the associated
chiral symmetry.  In Section~\ref{sec:spectral} we formulate the global
spectral problem and derive the purely damped relaxation modes.  In
Section~\ref{sec:green} we construct the retarded Green function,
susceptibility, and static response.  In
Section~\ref{sec:open_system} we integrate out the interior and obtain
the corresponding boundary effective theory.  We conclude in
Section~\ref{sec:discussion}.

We work in units where \(8\pi G=c=1\).

\section{Theory and background geometry}
\label{sec:background}

In this Section, building on \cite{Tasinato:2022vop,Atkins:2023axs}, we review the spherically symmetric,
asymptotically flat space-time solution which forms the 
 basis of our  subsequent analysis.
 We consider a non-minimally coupled, ghost-free  vector tensor theory defined by the Lagrangian density
\begin{equation}
\mathcal{L}
=
\frac{{\cal R}}{2}
-\frac{1}{4}F_{\mu\nu}F^{\mu\nu}
+\frac{1}{4}G_{\mu\nu}V^\mu V^\nu ,
\qquad
F_{\mu\nu}
=
\partial_\mu V_\nu-\partial_\nu V_\mu ,
\label{lagrangian}
\end{equation}
where ${\cal R}$ is the Ricci scalar, $G_{\mu\nu}$ is the Einstein tensor,
and $V^\mu$ is a vector field.
The non-minimal coupling $G_{\mu\nu}V^\mu V^\nu$ breaks the Abelian
gauge symmetry of the Maxwell term, and renders the longitudinal
component of the vector field dynamically relevant.

\smallskip

We look for static, spherically symmetric solutions of the form
\begin{equation}
\dd s^2
=
-A(r)\,\dd t^2
+\frac{\dd r^2}{B(r)}
+r^2\dd\Omega_2^2 ,
\qquad
V_\mu\,\dd x^\mu
=
\alpha_0(r)\,\dd t+\Pi(r)\,\dd r ,
\label{ansatz}
\end{equation}
where $\dd\Omega_2^2$ is the metric on the unit two-sphere.
The field equations admit two branches of solutions, depending
on whether the function $\Pi(r)$ vanishes, or not. One branch of solutions
corresponds to $\Pi(r)=0$, and it is continuously connected to the  Reissner-Nordstr\"{o}m configuration of General Relativity. 
The other branch with a non-vanishing $\Pi$ profile
is
 disconnected from the first, and leads
 to a Schwarzschild solution outside the object we consider -- hence it
 resembles the behavior of General Relativity in  vacuum. 
   Since this is the context 
 we wish to explore, we  focus on this branch in what follows. Nevertheless, depending on boundary conditions, both branches of solutions can coexist.   
The equations allow one to solve algebraically for the metric function $B$ and 
 the radial vector component $\Pi$. These quantities satisfy~\cite{Tasinato:2022vop} 
\begin{equation}
B(r)
=
\frac{A(r)}{A(r)+r A'(r)} \hskip0.5cm ,\hskip0.5cm
\label{B_from_A}
\Pi^2(r)
=
\frac{r\alpha_0^2}{A}
\left[
\frac{A'}{A}
-2\frac{\alpha_0'}{\alpha_0}
-r\frac{\alpha_0'^2}{\alpha_0^2}
\right] .
\end{equation}
We note that the reality of the radial profile requires $\Pi^2(r)\geq
0$, which imposes a non-trivial constraint on the allowed parameter
space. 

The temporal component $\alpha_0(r)$ and the metric function $A(r)$
satisfy two independent coupled differential equations,
\begin{equation}
0
=
\alpha_0''
+
\left[
\frac{2}{r}
-
\frac{2A'+rA''}{2\left(A+rA'\right)}
\right]\alpha_0'
+
\left[
\frac{A^{\prime 2}}{2A^2}
-
\frac{2A'+rA''}{2Ar}
\right]\alpha_0 \,,
\label{eq:reduced_alpha_equation}
\end{equation}
and
\begin{align}
0
={}&
\frac{\mathrm d}{\mathrm dr}
\Bigg\{
\left(
\frac{A}{A+rA'}
\right)^{3/2}
\left[
\frac{r\alpha_0^2 A'}{A^2}
-
\frac{2r\alpha_0\alpha_0'}{A}
-
\frac{r^2\alpha_0^{\prime 2}}{A}
\right]
+
\left(
\frac{A}{A+rA'}
\right)^{1/2}
\left[
1-\frac{\alpha_0^2}{4A}
\right]
\Bigg\}
\nonumber\\
&+
\left(
\frac{A}{A+rA'}
\right)^{1/2}
\frac{\mathrm d}{\mathrm dr}
\left(
\frac{\alpha_0^2}{4A}
\right)
+
\frac{r\alpha_0^{\prime 2}}{A}
\left(
\frac{A}{A+rA'}
\right)^{1/2}
\nonumber\\
&-
\frac{A'}{2\left(A+rA'\right)}
\left(
\frac{A}{A+rA'}
\right)^{1/2}
\left[
\frac{r\alpha_0^2 A'}{A^2}
-
\frac{2r\alpha_0\alpha_0'}{A}
-
\frac{r^2\alpha_0^{\prime 2}}{A}
\right]
\nonumber\\
&-
\frac{A'}{A+rA'}
\left(
\frac{A+rA'}{A}
\right)^{1/2}
\left[
2+\frac{\alpha_0^2}{2A}
\right] \,.
\label{eq:reduced_A_equation}
\end{align}
These equations admit a solution
   describing at once the exterior and the interior of a gravitationally
 bound,  asymptotically flat, and horizonless compact object: \cite{Tasinato:2022vop}. 
 We review the solution here, since it is the configuration
around which we are going to study the dynamics of fluctuations.

\subsection{Interior geometry}

The interior region $0 \leq r \leq R$ is described by the
configuration
\begin{align}
\alpha_0^{\rm int}(r)
&=
2Q_I\frac{R}{r}
+2\sigma
+\frac{2(1-\sigma)}{1+\gamma}
\left(\frac{r}{R}\right)^\gamma ,
\label{alpha_int}
\\
A_{\rm int}(r)
&=
\sigma^2
+\frac{2\sigma(1-\sigma)}{1+\gamma}
\left(\frac{r}{R}\right)^\gamma
+\frac{(1-\sigma)^2}{1+2\gamma}
\left(\frac{r}{R}\right)^{2\gamma} ,
\label{A_int}
\\
B_{\rm int}(r)
&=
\frac{A_{\rm int}(r)}
{\left[\sigma+(1-\sigma)(r/R)^\gamma\right]^2} ,
\label{B_int}
\end{align}
where $R$ is the radius of the compact object, $\sigma$ and $\gamma$
are dimensionless parameters, and $Q_I$ is an interior vector charge.
The limit $\sigma \to 0$ recovers a singular isothermal sphere profile \cite{Shu:1977uc}.
Regularity of the geometry in the interior 
 requires
\begin{equation}
\sigma \neq 0 ,
\end{equation}
which ensures
\begin{equation}
A_{\rm int}(0) = \sigma^2 > 0 ,
\qquad
B_{\rm int}(0) = 1 .
\end{equation}
These are the standard conditions for a regular center in a
spherically symmetric spacetime. Note that metric regularity at the
origin does not by itself constrain the vector potential $\alpha_0$
to be finite there.

To assess curvature  regularity, we compute the Ricci scalar ${\cal R}_{\rm int}$ of the
interior solution. Introducing the dimensionless coordinate $x \equiv
r/R$, we find
\begin{align}
{\cal R}_{\rm int}
&=
\frac{2\gamma(1-\sigma)}{R^2}
\frac{x^{\gamma-2}}
{\left[\sigma+(1-\sigma)x^\gamma\right]^3}
\nonumber\\
&\quad\times
\left[
\sigma^2
+\frac{\sigma(2-\gamma)(1-\sigma)}{1+\gamma}x^\gamma
+\frac{(1-\gamma)(1-\sigma)^2}{1+2\gamma}x^{2\gamma}
\right].
\end{align}
For $\sigma > 0$ the Ricci scalar is finite at $r = 0$ provided
$\gamma \geq 2 $.
 Other independent curvature invariants behave analogously under the
same condition. Also the energy momentum tensor -- which is anisotropic and
reads 
\be
\label{exp_EMTg}
T^{\mu}_{\,\,\nu}\,=\,{\rm{diag}}\left[ -\frac{{A}(r) \left(r {A}''(r)+2 {A}'(r)\right)}{r \left(r {A}'(r)+{A}(r)\right)^2},0,\frac{A'(r) \left(r A''(r)+2 A'(r)\right)}{4 \left(r A'(r)+A(r)\right)^2}, \frac{A'(r) \left(r A''(r)+2 A'(r)\right)}{4 \left(r A'(r)+A(r)\right)^2}\right]
\ee
is regular at the origin, under the same  conditions. 
 Throughout this paper we therefore assume $\sigma > 0$
and $\gamma \geq 2$, which defines the  interior branch for the geometry.

 \smallskip

The reality condition $\Pi^2_{\rm int}(r) \geq 0$ imposes an
additional constraint on the parameters. For the singular interior
$\sigma = 0$ this reduces to the simple inequality
\begin{equation}
Q_I
\geq
\frac{1}{\sqrt{1+2\gamma}}
-
\frac{1}{1+\gamma} \,
\ge 0.
\end{equation}
From now on, we assume that in general
 $Q_I\ge 0$. 
For the  branch $\sigma \neq 0$ the corresponding condition is
less compact and must be imposed as
\begin{equation}
\min_{0 \leq r \leq R} \Pi^2_{\rm int}(r) \geq 0 ,
\end{equation}
which defines an allowed region in the $(\sigma, \gamma, Q_I)$
parameter space.

\subsection{Exterior Schwarzschild branch}

The exterior region $r > R$ is described by the Schwarzschild metric,
\begin{equation}
\dd s^2_{\rm ext}
=
-\left(1-\frac{2M}{r}\right)\dd t^2
+\frac{\dd r^2}{1-2M/r}
+r^2\dd\Omega_2^2 ,
\end{equation}
with the vector field given by
\begin{align}
\alpha_0^{\rm ext}(r)
&=
2+2Q_E\frac{R}{r} ,
\label{alpha_ext}
\\
\Pi_{\rm ext}^2(r)
&=
\frac{4Q_E^2R^2+8Mr+8Q_ERr}{(r-2M)^2} .
\label{Pi_ext}
\end{align}
Here $M$ is the ADM mass and $Q_E$ is an exterior vector charge.
In order to ensure $\Pi_{\rm ext}^2(r)\ge0$, from now on
we set $Q_E\ge0$.
Notably,  while the metric is exactly Schwarzschild,
the vector profile $(\alpha_0^{\rm ext}, \Pi_{\rm ext})$ is
not trivial and differs from the Schwarzschild vacuum.
This distinction is physically important: odd-parity perturbations
couple to the vector background, and therefore the perturbative
response of this object differs from that of a Schwarzschild black
hole, even though the two spacetimes share the same exterior geometry.

\subsection{Matching conditions and compactness}
\label{sec_compt}

The compact object is constructed by smoothly joining the interior
solution to the exterior Schwarzschild branch at the surface $r = R$.
Since neither region contains a localized energy-momentum source at
the surface, the matching conditions are simply continuity of the
metric and of the vector field, as well as their derivatives at $r=R$.
 No distributional stress tensor is required by the junction conditions: no membrane is introduced.
The surface $r = R$ is simply the locus where the two analytic
branches of the solution effortlessly coincide.

Imposing Israel junction conditions  on the interior and exterior 
solutions described above determines the mass and
exterior charge in terms of the interior parameters:
\begin{equation}
M
=
\frac{(1-\sigma)\gamma(1+\gamma+\sigma\gamma)}
{(1+\gamma)(1+2\gamma)}
R ,
\qquad
Q_E
=
Q_I
-
\frac{\gamma(1-\sigma)}{1+\gamma} \,,
\label{mass_charge}
\end{equation}
and recall that in what follows we assume $Q_E\ge0$. Hence, 
$A$, $A'$, and $B$ match at $r=R$, and the vector profile is matched by choosing the exterior charge $Q_E$ as in Eq.~\eqref{mass_charge}. The surface $r=R$ should therefore be understood as a regular matching surface with no membrane source, although not all radial derivatives of the metric functions need be continuous there.

The compactness of the object is therefore
\begin{equation}
\mathcal{C}
\equiv
\frac{M}{R}
=
\frac{(1-\sigma)\gamma(1+\gamma+\sigma\gamma)}
{(1+\gamma)(1+2\gamma)} .
\label{compactness}
\end{equation}

\begin{figure}[t]
\vspace{-10pt}
\centering
\includegraphics[width=0.5\textwidth]{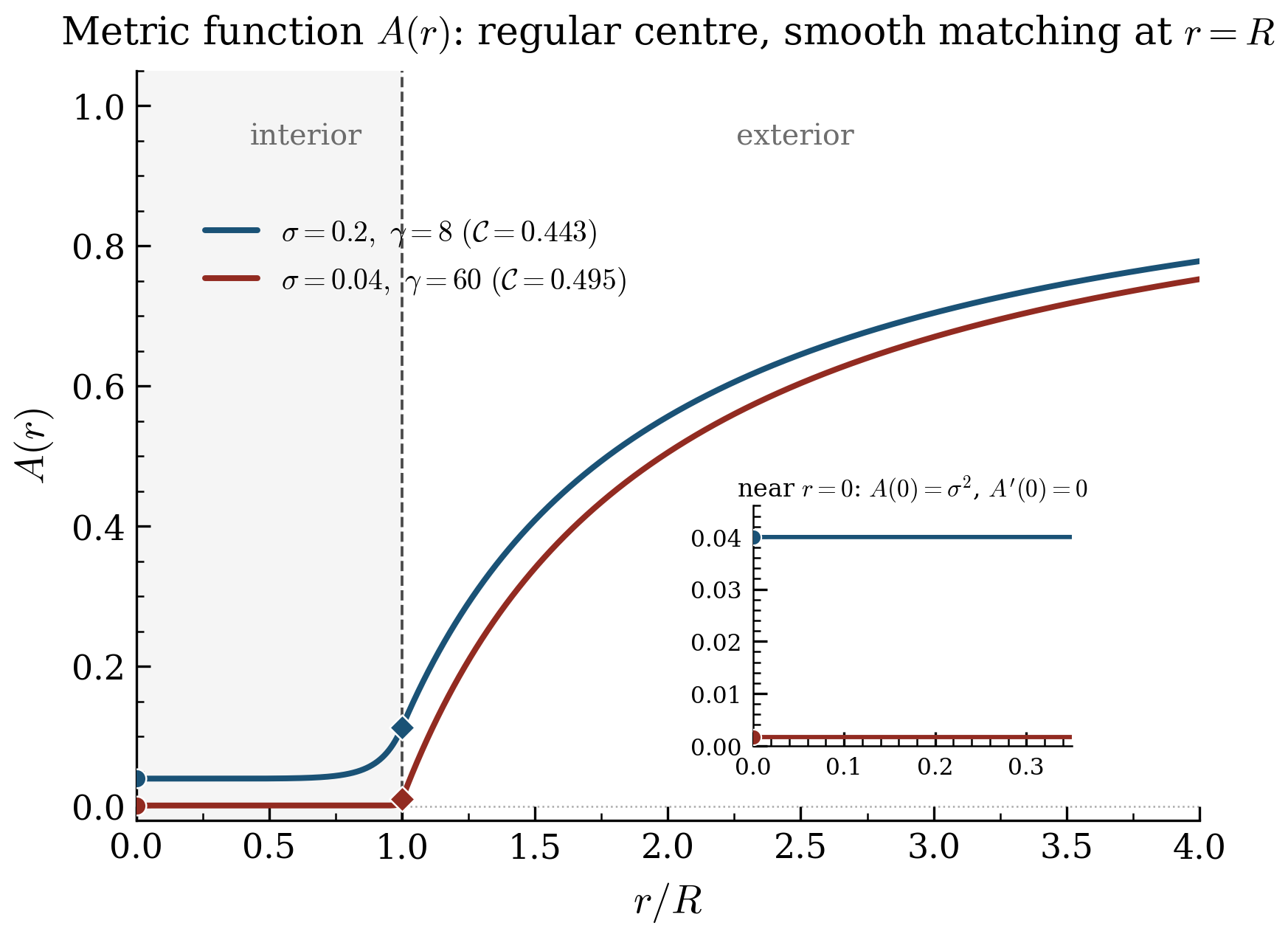}
\includegraphics[width=0.49\textwidth]{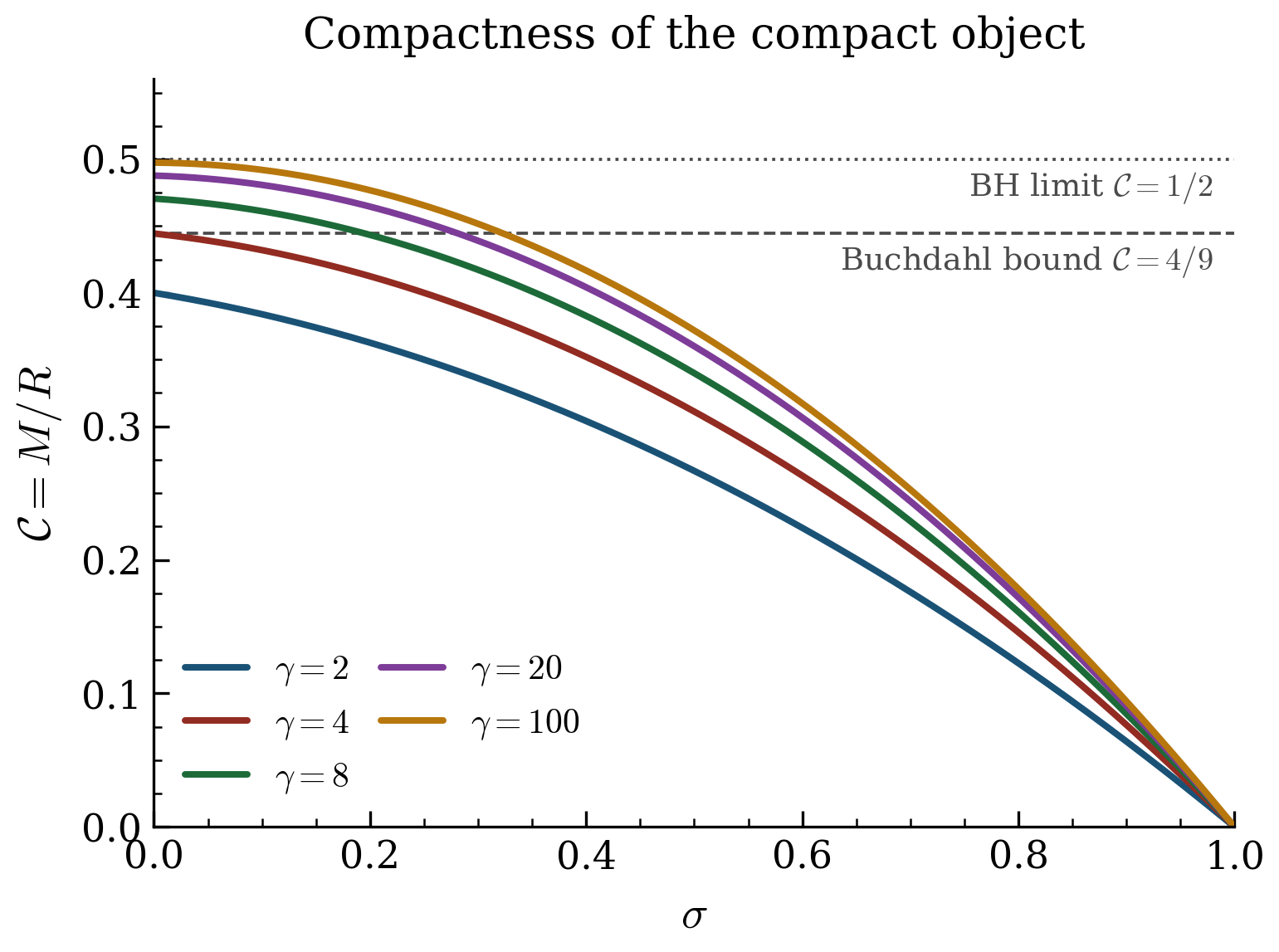}
\caption{\small \it {\bf Left:}
Metric function $A(r)$ across the regular interior ($r<R$, shaded) and the Schwarzschild exterior ($r>R$), for a moderate-compactness configuration and a near-black-hole configuration approaching $\mathcal{C}\to1/2$ via Eq.~\eqref{bh_limit}. In both cases $A(r)$ is finite and flat at the origin, $A(0)=\sigma^2$ with $A'(0)=0$ (circles; see inset), and joins continuously, with matched slope, onto the exterior branch at $r=R$ (diamonds): the interior-exterior matching is smooth, with no membrane or kink at the surface.
 {\bf Right:}  Compactness $\mathcal{C}=M/R$ of the interior solution as a function of the regularity parameter $\sigma$, for several values of the vector-coupling parameter $\gamma$, Eq.~\eqref{compactness}. The dashed line marks the Buchdahl bound $\mathcal{C}=4/9$, which can be violated here thanks to the anisotropic stress sourced by the vector profile; the dotted line marks the black-hole value $\mathcal{C}=1/2$, approached in the double limit $\sigma\to0$, $\gamma\to\infty$, Eq.~\eqref{bh_limit}.}
\label{fig_metrfunc}
\end{figure}

 The configuration can avoid the Buchdahl bound ${\cal C}\le 4/9$ thanks to 
a large anisotropic stress induced by the vector field profile \cite{Tasinato:2022vop}
(see also \cite{Raposo:2018rjn}),
see Eq.~\eqref{exp_EMTg}. See
Fig.~\ref{fig_metrfunc}, right panel.
  In fact,
the compactness approaches the black-hole value
$\mathcal{C} = 1/2$  in the  double limit
\begin{equation}
\sigma \to 0 ,
\qquad
\gamma \to \infty .
\label{bh_limit}
\end{equation}
The physical significance of this limit is discussed in
Section~\ref{sec:spectral}, where we show that it corresponds to a special 
interior-exterior decoupling limit.

We emphasize that the object constructed above has no membrane.
The matching surface $r = R$ carries no localized stress-energy, and
the background solution is globally smooth. See
Fig.~\ref{fig_metrfunc}, left panel, for a representation of the smooth metric
function $A(r)$.
The spectral problem studied in the following sections is therefore
not a membrane or thin-shell problem, but a  global problem
for a smooth  compact object with a regular geometry. In this
sense, our setup is conceptually distinct
from the physics of exotic compact objects
that are typically equipped with  a dynamical
surface \cite{Thorne:1986iy}. The corresponding parity-odd  
 fluctuations have several intriguing properties, as we are going to
  learn, and make the system very different, hence distinguishable, from
  configurations in General Relativity with similar exterior geometries.

\section{Odd-parity perturbations and underlying symmetry}
\label{sec:perturbations}

We analyze the dynamics of fluctuations around the spherically symmetric configuration
of the previous section. We focus on the odd-parity modes, since they are the simplest
to study, and since they directly track the properties of the vector degrees of freedom
characterizing  the  system anisotropic stress. Remarkably, we will find 
that the dynamics of fluctuations is  fully analytically  tractable, and 
is characterized by an underlying chiral symmetry
that determines its properties.

\smallskip

The odd-parity perturbation equations for both the spin-1 vector sector
and the spin-2 tensor sector have been  explored in \cite{Atkins:2023axs} for the exterior of the geometry: here
we extend the analysis to the interior of the compact space-time. 
They 
 are governed by two background functions
that can be constructed from the metric and vector field of
Section~\ref{sec:background}. We define
\begin{equation}
\Delta_1(r)
=
\frac{r\,\alpha_0(r)}
{2\sqrt{A(r)+r A'(r)}} \,\,\hskip0.5cm,
\label{def_Delta1}
\hskip0.5cm
\Sigma_1(r)
=
\frac{r\,\sqrt{\alpha_0^2(r)-4A(r)}}{2A(r)} .
\end{equation}
These functions are evaluated with the appropriate background --- interior
or exterior --- in each region.
Both appear in the first-order transport operators that emerge from the
factorization properties described in what follows.
For later use we record the value of $\Sigma_1$ at the origin:
using the interior solution~\eqref{alpha_int}--\eqref{A_int} and
recalling that $\alpha_0^{\rm int}(r) \sim 2Q_I R/r$ as $r \to 0$ 
we find
\begin{equation}
\Sigma_1(0)
=
\frac{Q_I R}{\sigma^2} \,\equiv\,\Sigma_0.
\label{Sigma1_origin}
\end{equation}
The quantity $\Sigma_0$ plays a central role in characterizing  the 
spectrum of fluctuations of  the compact object. Notice that it normally diverges
as $\sigma^2\to0$.

\subsection{Spin-1 axial vector perturbation}

We decompose the odd-parity vector perturbation in terms of
vector spherical harmonics $ Y_{\ell m}(\theta,\phi)$:
\begin{equation}
\delta V_\mu\,\dd x^\mu
=
\frac{a(t,r)}{\sin\theta}
\left( \partial_\phi Y_{\ell m} \right)\,\dd\theta
-a(t,r)\sin\theta\,\left(\partial_\theta Y_{\ell m}\right)\,\dd\phi ,
\end{equation}
and introduce the normalized master variable $\beta(t,r)$ by
\begin{equation}
a(t,r)
=
\frac{r\,\alpha_0^2(r)}{4}\,\beta(t,r) .
\end{equation}
The perturbation equations then reduce to a single master equation
for $\beta$, valid in both the interior and the exterior,
\begin{align}
0
&=
\Sigma_1^2\,\partial_t^2\beta
-2\Sigma_1\Delta_1\,\partial_t\partial_r\beta
+\Delta_1^2\,\partial_r^2\beta
\nonumber\\
&\quad
-\left(\Sigma_1'\Delta_1+\Sigma_1\right)\partial_t\beta
+\Delta_1\left(\Delta_1'+1\right)\partial_r\beta
-\ell(\ell+1)\beta .
\nonumber\\
&\equiv \mathcal{E}^{(1)}_\ell \left[ \beta \right]
\label{spin1_master}
\end{align}
We restrict to $\ell \ge 1$. The structure of this equation is not that of a standard wave equation
in a potential. It admits a factorization into commuting first-order
operators 
\begin{equation}
\label{timed_dj1}
{\cal D}^{(1)}_{\pm} \,=\,\mathcal{J}_t-\lambda^{(1)}_{\pm}
\end{equation}
where $\mathcal{J}_t$ is the  vector field
\begin{equation}
\label{def_jt1}
\mathcal{J}_t
= \Delta_1(r)\partial_r-
\Sigma_1(r)\partial_t ,
\end{equation}
while
 the constants $\lambda^{(1)}_\pm$ are
\begin{equation}
\lambda^{(1)}_+ = \ell \quad ,
\quad
\lambda^{(1)}_- = -(\ell+1) .
\label{exp_lam1}
\end{equation}
It is straightforward to 
 verify that
\begin{equation}
\mathcal{E}^{(1)}_\ell
\equiv
\mathcal{D}^{(1)}_-\mathcal{D}^{(1)}_+
=
\mathcal{D}^{(1)}_+\mathcal{D}^{(1)}_- .
\label{factorization_spin1}
\end{equation}
 The commutativity of the two operators is a non-trivial property of
the background functions $\Delta_1$ and $\Sigma_1$, and underlies
the exact solvability of the perturbation problem. It is also related with an underlying
chiral symmetry of the system, a topic that 
we cover in Section \ref{subsec:quadratic_action_correct}.

\subsection{Spin-2 axial metric perturbation}

In Regge--Wheeler gauge, the odd-parity metric perturbation is
parametrized by two functions $h_0(t,r)$ and $h_1(t,r)$,
\begin{equation}
h_{\mu\nu}^{\rm odd}
=
\begin{pmatrix}
0 & 0 & -h_0\,\partial_\phi Y/\sin\theta
      & h_0\sin\theta\,\partial_\theta Y
\\
0 & 0 & -h_1\,\partial_\phi Y/\sin\theta
      & h_1\sin\theta\,\partial_\theta Y
\\
-h_0\,\partial_\phi Y/\sin\theta
& -h_1\,\partial_\phi Y/\sin\theta
& 0 & 0
\\
h_0\sin\theta\,\partial_\theta Y
& h_1\sin\theta\,\partial_\theta Y
& 0 & 0
\end{pmatrix}.
\end{equation}
The function $h_1$ satisfies an algebraic equation and can be
eliminated in terms of the remaining variables~\cite{Atkins:2023axs}.
Introducing the rescaled master variable $\Gamma$ as 
\begin{equation}
h_0(t,r) = r^2\,\Gamma(t,r) ,
\end{equation}
the resulting spin-2 parity-odd equation takes the form (we restrict to $\ell \ge 2$)
\begin{align}
F_\ell[\beta]
&=
\Sigma_1^2\,\partial_t^2\Gamma
-2\Sigma_1\Delta_1\,\partial_t\partial_r\Gamma
+\Delta_1^2\,\partial_r^2\Gamma
\nonumber\\
&\quad
-\left(\Sigma_1'\Delta_1+3\Sigma_1\right)\partial_t\Gamma
+\Delta_1\left(\Delta_1'+3\right)\partial_r\Gamma
-(\ell-1)(\ell+2)\Gamma ,
\label{spin2_master}
\end{align}
where $F_\ell[\beta]$ is a source term built from the spin-1
perturbation $\beta$ and its derivatives, which vanishes when
$\beta = 0$.
The homogeneous part of this equation, obtained by setting
$F_\ell[\beta] = 0$, also factorizes. Defining
\begin{equation}
\label{timed_dj}
{\cal D}^{(2)}_{\pm} \,=\,\mathcal{J}_t-\lambda^{(2)}_{\pm}
\end{equation}
where $\mathcal{J}_t$ is given in Eq.~\eqref{def_jt1}
 and the constants $\lambda^{(2)}_\pm$ are
\begin{equation}
\label{lam_j2}
\lambda^{(2)}_+ = \ell -1 \quad,
\quad
\lambda^{(2)}_- = -(\ell+2) \,.
\end{equation}
We find
\begin{equation}
\mathcal{E}^{(2)}_\ell
\equiv
\mathcal{D}^{(2)}_-\mathcal{D}^{(2)}_+
=
\mathcal{D}^{(2)}_+\mathcal{D}^{(2)}_- ,
\label{factorization_spin2}
\end{equation}
so that the homogeneous spin-2 equation reads $\mathcal{E}^{(2)}_\ell
\Gamma = 0$. 
The factorized structure is identical in form to the spin-1 case,
with  a shift in the constants $\lambda^{(2)}_\pm$
 with respect to the $\lambda^{(1)}$'s of Eq.~\eqref{exp_lam1}.
Therefore the two sectors, in the homogeneous limit,  share the same characteristic geometry but
differ in the exponents governing their spatial profiles. As we will learn
in what follows, the equations
are simple enough to be solved exactly, and the corresponding
Green functions can be determined analytically -- see Section \ref{sec:green}. Having the exact Green function,
it is  straightforward to obtain a solution for the
inhomogeneous Eq.~\eqref{spin2_master} for the spin-2 sector.

\subsection{Characteristic structure and unidirectional transport}

The factorizability property of Eqs~\eqref{factorization_spin1}, \eqref{factorization_spin2} --
both in the exterior and in the interior of the geometry 
-- 
 implies that both second-order master equations are built from
commuting first-order transport operators sharing the same
characteristic direction. 
 We now briefly comment on some physical ramifications
of these findings, which will be further elaborated below.
  While the homogeneous master equation for the 
vector spin-1 sector  completely describes
the vector fluctuations, the master equation for the spin-2 sector -- as we learned --
has also a vector source. We focus in what follows on the sourceless,  homogeneous
limit of the spin-2 dynamics, since the effects of the vector source 
can be analyzed by means of  the Green function method.

\smallskip

\begin{figure}[t]
\centering
                \includegraphics[width=0.5\linewidth]{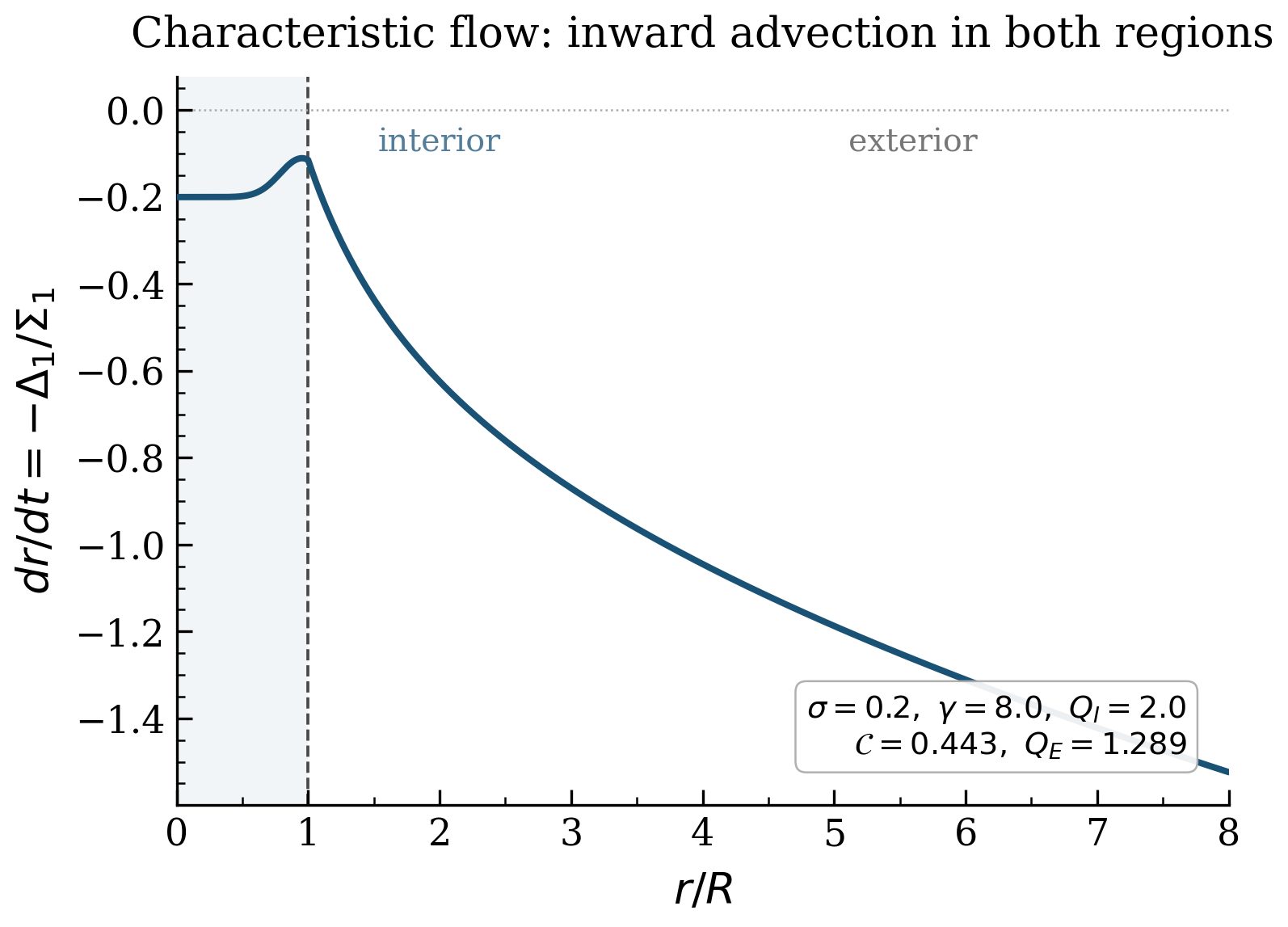}
\caption{\small \it Characteristic speed $dr/dt=-\Delta_1/\Sigma_1$ of the odd-parity transport equation, shown across the interior ($r<R$, shaded) and exterior ($r>R$) for a representative configuration ($\sigma=0.20$, $\gamma=8$, $Q_I=2$). The flow is everywhere negative, showing that characteristics propagate inward in both regions: the odd sector is a unidirectional advection problem rather than a two-way wave equation, with no characteristic available to carry energy back out to infinity. }
\label{fig:AAa1}
\end{figure}

It is convenient to treat both sectors uniformly by writing
\begin{equation}
\Phi_j
=
\begin{cases}
\beta , & j = 1 , \\
\Gamma , & j = 2 ,
\end{cases}
\end{equation}
so that the master equations take the unified form
$\mathcal{E}^{(j)}_\ell\Phi_j = 0$,
The first order operators ${\cal D}^{(j)}_{\pm} $ whose product  forms $\mathcal{E}_\ell^{(j)}$
are given by Eqs~\eqref{timed_dj1},~\eqref{timed_dj} respectively. 
The  integral curves associated with the vector flux field $\mathcal{J}_t$ satisfy
\begin{equation}
\frac{\dd r}{\dd t}
=
-\frac{\Delta_1(r)}{\Sigma_1(r)} .
\label{characteristics}
\end{equation}
When $\Delta_1/\Sigma_1 > 0$, as it happens in our setup,
see Fig.~\ref{fig:AAa1},  perturbations are advected toward
smaller $r$, i.e.\ toward the interior of the object.

The appropriate characteristic coordinate adapted to the flow is then
\begin{equation}
\nu(t,r) = t + \chi(r) ,
\qquad
\chi(r) = \int^r_R \frac{\Sigma_1(\rho)}{\Delta_1(\rho)}\,\dd\rho ,
\label{def_nu}
\end{equation}
while a natural radial variable for this system is
\begin{equation}
y(r) = \int^r_R \frac{\dd\rho}{\Delta_1(\rho)} .
\label{def_y}
\end{equation}
The coordinate $\chi(r)$ plays the role of a tortoise coordinate adapted
to our setup, in analogy with the standard tortoise
coordinate of black-hole perturbation theory. With this convention \(y(R)=0\). Since \(\Delta_1>0\) for the solution we
consider, we have \(y<0\) in the interior and in particular
\(y_0\equiv y(0)<0\), while \(y>0\) in the exterior.

\smallskip

In terms of the quantities $\left(\nu(t,r),\, y(r)\right)$, the general  solution of either master
equation is  
\begin{equation}
\Phi_j(\nu, y)
=
F_+(\nu)\,e^{\lambda^{(j)}_+ y}
+
F_-(\nu)\,e^{\lambda^{(j)}_- y} ,
\label{general_solution}
\end{equation}
as a linear combination of the solution of first-order equations ${\cal D}^{(j)}_{\pm} \Phi_j=0$.
The quantities $F_\pm$ in Eq.~\eqref{general_solution}  are arbitrary functions of $\nu(t,r)$ alone, and the
exponents are controlled by the quantities $\lambda^{(j)}_\pm$ of Eqs.~\eqref{exp_lam1} and \eqref{lam_j2}.

For harmonic modes $\Phi_j \propto e^{-i\omega t}$, the solution
becomes
\begin{equation}
\Phi_j(t,r)\,=\, e^{-i\omega t}\,\psi_j(r)
\label{harmonic_solution_a}
\end{equation}
with
\begin{equation}
\psi_j(r) =
e^{-i\omega \chi(r)}
\left[
C^{(j)}_+ e^{\lambda^{(j)}_+ y}
+
C^{(j)}_- e^{\lambda^{(j)}_- y}
\right] .
\label{harmonic_solution}
\end{equation}
In other words, after passing to Fourier space and introducing the first-order operator
\begin{equation}
\mathcal{J}_\omega
= \Delta_1(r)\partial_r+i \omega
\Sigma_1(r),
\label{def_jome}
\end{equation}
 the quantities entering  \eqref{harmonic_solution} are solutions of the first order equations
 \begin{equation}
 \left(\mathcal{J}_\omega-
 \lambda_{\pm}^{(j)}\right)\,\psi_j\,=\,0
 \end{equation}
 corresponding to the Fourier version of the time-domain
 first order operators \eqref{timed_dj}.
A key property is that both contributions in~\eqref{harmonic_solution_a}, \eqref{harmonic_solution}
depend on time through the same combination $e^{-i\omega(t+\chi(r))}$.
This is fundamentally different from the standard Regge--Wheeler or
Zerilli problem, where the two independent solutions carry
$e^{-i\omega(t\pm r_\star)}$ with opposite signs in the tortoise
coordinate $r_\star$ --  corresponding to ingoing and outgoing wave
branches.
In the present case there is {\it only one} characteristic direction: both
branches of the solution are advected along the same null-like combination
$\nu$.
The system therefore supports unidirectional transport rather than
two-way wave propagation, as  in ordinary black-hole perturbation theory.

\smallskip

We emphasize that this unidirectionality does not imply that the
system continuously emits radiation from infinity.
Rather, it means that initial data and localized sources are advected along the
preferred flow defined by the background vector field, and the
dynamics is that of an unidirectional,  chiral transport medium rather than a
conventional wave equation.
This property, as we will learn, has important  consequences for the intrinsic  spectral properties of the system, and for the response functions of the object to external perturbations.

\subsection{Quadratic action and an underlying chiral symmetry}
\label{subsec:quadratic_action_correct}

We  derive a quadratic variational principle for the homogeneous
factorized equations discussed above.  This  action  allows us to identify
  the  boundary flux associated with the transport of fluctuations, 
which turns out to  be useful in Section \ref{sec:spectral}. It also 
 unravels  underlying chiral (and conformal) symmetries for the system.
  The construction applies  to the spin-1 sector and
to the homogeneous part of the spin-2 sector.

\smallskip

Owing to the factorizability of the homogeneous evolution equations, we recast them as
\begin{equation}
\left(\mathcal{J}_t-\lambda^+_j \right)\left( \mathcal{J}_t-\lambda^-_j \right)\,\Phi_j =0 
\hskip0.5cm\Rightarrow\hskip0.5cm
\left[
\mathcal{J}_t^2
-
s_j \mathcal{J}_t
-
\Lambda_j
\right]\Phi_j =0 ,
\label{eq:homogeneous_transport_equation}
\end{equation}
with 
\begin{equation}
s_j = \lambda_j^{(+)}+\lambda_j^{(-)}\,,
\hskip0.7cm  \Lambda_j
\,=\,- \lambda_j^{(+)} \,\lambda_j^{(-)}\ge0\,.
\label{def_olam}
\end{equation}
The first order operator $\mathcal{J}_t$ is defined in Eq.~\eqref{def_jt1}   while the coefficients $\lambda^{(j)}_\pm$
are   in Eqs.~\eqref{exp_lam1} and \eqref{lam_j2}.

\smallskip
We look for a quadratic action for parity-odd fluctuations 
 of the form
\begin{equation}
S_j^{(2)}
=
\frac{1}{2}
\int \dd t\,\dd r\,\mu_j(r)
\left[
\left(\mathcal{J}_t\Phi_j\right)^2
+
\Lambda_j \Phi_j^2
\right] .
\label{eq:quadratic_action_correct}
\end{equation}
for an appropriate measure $\mu_j(r)$,
whose variation gives the evolution 
equation \eqref{eq:homogeneous_transport_equation}.
By choosing
\begin{equation}
\mu_j(r)
=
\frac{1}{\Delta_1(r)}
\exp\!\left[
-s_j\int^r \frac{\dd \rho}{\Delta_1(\rho)}
\right]
\label{eq:correct_measure}
\end{equation}
when varying 
\eqref{eq:quadratic_action_correct} gives
\begin{equation}
\delta S_j^{(2)}
=
-\int \dd t\,\dd r\,\mu_j
\left[
\left(\mathcal{J}_t^2-s_j\mathcal{J}_t-\Lambda_j\right)\Phi_j
\right]\delta\Phi_j
+
\delta S_{j,\rm bdry}^{(2)} .
\end{equation}
as desired, with 
\begin{equation}
\delta S_{j,\rm bdry}^{(2)}
=
\int \dd t\,
\mu_j(r)\Delta_1(r)
\mathcal{J}_t[\Phi_j]\,
\delta\Phi_j
\bigg|_{\rm bdry}.
\label{eq:boundary_variation_correct}
\end{equation}
Hence action \eqref{eq:quadratic_action_correct} is the correct quadratic action giving rise to the homogeneous
equations for spin-1 and spin-2 fluctuations. Our procedure hence identifies a convenient measure
$\mu_j(r)$ for the space of functions on which the flux operator $\mathcal{J}_t$ 
acts which will turn out to be useful for determining the response of the object to external
perturbations. Also, Eq.~\eqref{eq:boundary_variation_correct} makes the boundary contributions
manifest -- an important piece of information for identifying the correct
matching conditions at the object surface.

\smallskip

The simple quadratic action \eqref{eq:quadratic_action_correct} can be recast in  a way to exhibit
interesting symmetries. 
Changing the variables $(t,r)$ to the characteristic variables $(\nu,y)$ defined in Eqs~\eqref{def_nu}, \eqref{def_y}, we have $\partial_t=\partial_\nu$,
 $
\partial_r
=
{\Sigma_1}/{\Delta_1}\,\partial_\nu
+
{1}/{\Delta_1}\,\partial_y $
 and therefore $\mathcal J_t=\partial_y$. The quadratic action reads then
\begin{equation}
S_j^{(2)}
=
\frac{1}{2}
\int d\nu\,dy\,
e^{-s_jy}
\left[
\left(\partial_y\Phi_j\right)^2
+
\Lambda_j\Phi_j^2
\right].
\label{weighted_action_sec8}
\end{equation}
The weight in Eq.~\eqref{weighted_action_sec8} can be removed from the
bulk action by the field redefinition
$\Phi_j(\nu,y)
=
e^{s_jy/2}\Psi_j(\nu,y)
$, finding
\begin{align}
S_j^{(2)}
&=
\frac{1}{2}
\int d\nu\,dy
\left[
\left(\partial_y\Psi_j\right)^2
+
\left(
\ell+\frac12
\right)^2\Psi_j^2
\right]
+
\frac{s_j}{4}
\int d\nu\,dy\,\partial_y\left(\Psi_j^2\right).
\label{reduced_bulk_action_sec8}
\end{align}
The last term is a boundary term. Importantly, this action does not contain derivatives
along the variable $\nu$ -- this is the key property underlying the symmetry.  Its corresponding Euler-Lagrange equations are easily solvable, and lead to the solution \eqref{general_solution}, once expressed in terms of the original quantity $\Phi_j(\nu, y)$.

\paragraph{Chiral symmetry.}
The form of Eq.~\eqref{reduced_bulk_action_sec8} makes a chiral symmetry
manifest. In fact, the action  is  invariant, up to a total derivative, under the
infinitesimal transformation corresponding to a chiral symmetry
\begin{equation}
\delta_\epsilon \Psi_j
=
\epsilon(\nu)\partial_\nu\Psi_j
+
\frac{1}{2}\epsilon'(\nu)\Psi_j .
\label{chiral_transformation_sec8}
\end{equation}
While  Eq.~\eqref{reduced_bulk_action_sec8}   is invariant under a general chiral reparametrizations of $\nu$, a restricted,  global \(SL(2,{\rm R})\) symmetry is obtained by restricting
\(\epsilon(\nu)\) to be a quadratic polynomial,
\begin{equation}
\epsilon(\nu)=a+b\nu+c\nu^2 .
\end{equation}
Defining $\epsilon_n (\nu) = \nu^{n+1}$, we can write the previous expression
as $\epsilon=a \epsilon_{-1}+b\epsilon_{0}+c \epsilon_{1}$. Introducing the differential operators
\begin{equation}
L_n\,=\,
-\epsilon_n (\nu) \,\partial_\nu-\frac12\,\epsilon_n' (\nu)
\end{equation}
as generators of the symmetry, we find they
 satisfy the  \(SL(2,{\rm R})\) conformal algebra
\begin{equation}
[L_m,L_n]\,=\,(m-n)L_{m+n} .
\label{Ln_algebra_sec8}
\end{equation}
The chiral (or, in a special case,  conformal) symmetry discussed
   reflects  the unidirectional transport of the system. In
   fact, the factorized operators ${\cal D}_\pm$ for the evolution equations that
   we met in Section~\ref{sec:perturbations} are related with the generators
   of the conformal symmetry \cite{Atkins:2023axs}.
  Such a symmetry is valid throughout the space-time -- not only in proximity
  of the object surface. The study of underlying and hidden
  symmetries characterizing the dynamics of fluctuations in black
  hole space-times have a long history and interesting ramifications,
  see e.g. \cite{Carlip:1998wz,Solodukhin:1998tc,Birmingham:2001qa,Castro:2010fd,Bertini:2011ga,Chen:2010ik} and they recently became important for understanding the physics of  
  black hole (vanishing) Love numbers (see e.g. \cite{Hui:2020xxx,Charalambous:2021mea,Charalambous:2021kcz,Hui:2021vcv,Charalambous:2022rre,BenAchour:2022uqo,Lupsasca:2025pnt,Parra-Martinez:2025bcu}). We find it intriguing 
  that  our compact object in modified gravity is characterized
  by such a general chiral symmetry in its odd-parity fluctuation sector. 
  While
valid in the entire bulk of the geometry, the chiral symmetry can be broken by appropriate boundary conditions. This
is interesting and important, since it can lead to a quantized spectrum of
fluctuations, as we are going to discuss below. 

\section{Spectral problem and relaxation modes}
\label{sec:spectral}

Having solved the local problem of determining the analytic solutions of the
odd-parity equations, we now turn to the global spectral problem. Our aim is to
identify the source-free fluctuations selected by physically appropriate
boundary conditions.
In standard black-hole perturbation theory, the quasinormal-mode spectrum is
obtained by imposing ingoing waves at the horizon and outgoing waves at
infinity. This logic does not apply here. The two branches in
Eq.~\eqref{harmonic_solution} are not independently propagating ingoing and
outgoing waves: they propagate along the same characteristic direction and
differ only by their radial profiles \(e^{\lambda_\pm^{(j)}y}\). The spectrum
therefore arises only after the regular interior and the boundary conditions are
included. These conditions break the chiral symmetry of the local equations and
select a set of intrinsic fluctuations consisting purely of relaxation modes.

\subsection{Global spectral problem for the  compact object}
\label{sec_bou}

We now impose convenient boundary conditions
for the fluctuations analyzed in Section \ref{sec:perturbations}, with
interesting physical ramifications. Recall that
the  spacetime is smooth, and the perturbations should satisfy
boundary conditions as: 
\begin{enumerate}
\item[(a)] continuity of $\Phi_j$ and of the flux ${\cal J}_t$ (see Eq.~\eqref{eq:boundary_variation_correct})  at the matching surface $r = R$;
\item[(b)]  an appropriate   condition for $\Phi_j$ at the origin $r = 0$;
\item[(c)] an appropriate boundary condition as $r \to \infty$.
\end{enumerate}
Even though the local propagation is one-way, such  conditions
together  can  impose  non-trivial  constraints on the dynamics, breaking the chiral symmetry of Section~\ref{subsec:quadratic_action_correct}. 
 Accordingly, we select physically reasonable conditions (a)-(c)
which lead to a quantized spectrum with no instabilities and interesting features. 

\smallskip

We now focus on harmonic modes. 
We write the Fourier version of the interior ($r\le R$) and exterior ($r\ge R$) solutions of the corresponding homogeneous equations in the form
\begin{align}
\psi_j^{\rm int}(r)
&=
e^{-i\omega\chi(r)}
\left[
A^{(j)}_+ e^{\lambda_+^{(j)} y(r)}
+
A^{(j)}_- e^{\lambda_-^{(j)} y(r)}
\right] ,
\label{psi_int}
\\
\psi_j^{\rm ext}(r)
&=
e^{-i\omega\chi(r)}
\left[
B^{(j)}_+ e^{\lambda_+^{(j)} y(r)}
+
B^{(j)}_- e^{\lambda_-^{(j)} y(r)}
\right] ,
\label{psi_ext}
\end{align}
where the coordinates $\chi(r)$ and $y(r)$ are evaluated with the
appropriate interior or exterior background functions. Since the 
structure of the problem is the same for $j=1,2$, {\it from now on we 
suppress the spin index $j$}, unless otherwise stated.

\smallskip

\paragraph{(a) Matching  conditions at the surface}
Since the background contains no membrane and no localized
stress-energy at $r = R$, the matching conditions we impose at the surface are
simply continuity of the function and of the flux
 (recall the analysis of the boundary terms in Section \ref{subsec:quadratic_action_correct})
\begin{equation}
\left[\psi\right]_R = 0 ,
\qquad
\left[\Delta_1\psi' + i\omega\Sigma_1\psi\right]_R = 0 .
\label{matching_pert}
\end{equation}
 For the 
   class of solutions considered here, these
conditions identify the interior and exterior amplitudes of Eqs~\eqref{psi_int} and \eqref{psi_ext},
\begin{equation}
\label{equa_intex}
        A_\pm = B_\pm \, .
\end{equation}

\paragraph{(b) Boundary condition at the origin}

The boundary condition at the origin constrains the ratio of the two interior
coefficients $A^{(j)}_\pm$ appearing in Eq.~\eqref{psi_int}. 
We adopt a Neumann-type prescription
\begin{equation}
\partial_r\psi\big|_{r \to 0} = 0 .
\label{origin_bc}
\end{equation}
It
imposes, using \eqref{equa_intex}, 
\begin{equation}
\label{cond_impor}
0=
B_-\, e^{-(2\ell+1)y_0}
\left(
\omega+\frac{i\lambda_-}{\Sigma_0}
\right)
+
B_+
\left(
\omega+\frac{i\lambda_+}{\Sigma_0}
\right) ,
\end{equation}
where $y(0)=y_0$, and $\Sigma_1(0)=\Sigma_0$. The quantities $\lambda_+>0$ and $\lambda_-<0$ are the two characteristic exponents of
the radial problem, whose values depend only on the multipole number and on
the spin sector, see Eqs.~\eqref{exp_lam1} and \eqref{lam_j2}. Notice that the  boundary condition
\eqref{cond_impor} explicitly depends
on frequency: it violates the chiral symmetry \eqref{chiral_transformation_sec8} and it breaks the local unidirectional
transport discussed in the previous section. In fact, Eq.~\eqref{origin_bc} is a simple and parameter-free way to break the symmetry, and it should be understood as a minimal symmetry-breaking prescription at the regular centre. It is not claimed to be unique. Its role is to select a centre-regular branch without introducing additional dimensionful data. More general Robin conditions correspond to different microscopic gluing data and shift the dissipative pole.

\paragraph{(c) Asymptotic boundary condition and passivity}

The asymptotic boundary condition far from the object requires particular care, as it differs 
from the standard outgoing-wave prescription we outlined above. It provides the last ingredient
we need to determine the quantized frequency spectrum. 

In the ordinary quasinormal-mode problem, spatial infinity is an
outflow boundary: radiation escapes to infinity, where we impose
outgoing conditions.
In the present system the characteristics point inward, so spatial
infinity instead behaves as an \emph{inflow} boundary.
A physically convenient prescription -- which we adopt here -- is therefore not to select
outgoing radiation, but to select the asymptotic channel compatible
with a \emph{passive retarded response}. The compact object should
respond to incoming perturbations without spontaneously generating
exponentially growing modes in time.  Moreover, the resulting solutions should
exponentially decay at large spatial distances from the object. 

\smallskip

\begin{figure}[t]
\centering
                \includegraphics[width=0.5\linewidth]{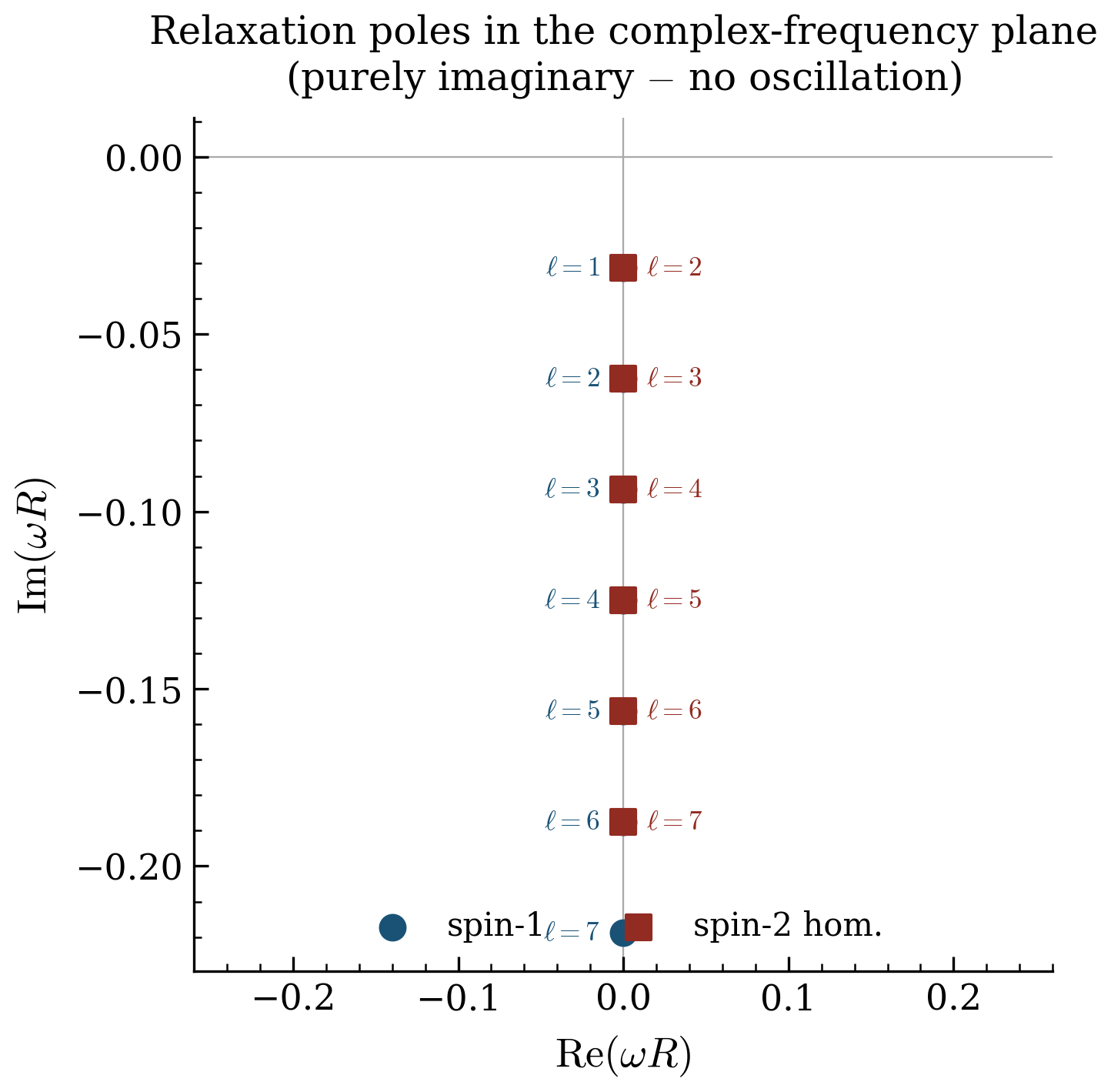}
\caption{\small \it Relaxation poles $\omega_\star$ in the complex-frequency plane for spin-1 (circles) and spin-2 homogeneous (squares) perturbations, at fixed $\sigma=0.25$, $Q_I=2$, labelled by multipole number $\ell$. All poles lie exactly on the negative imaginary axis, $\omega_\star=-i\Gamma_{\rm rel}^{(\ell)}$: the spectrum is purely damped, with no real part and hence no oscillatory ringing. The relaxation rate grows linearly with $\ell$. }
\label{fig:AA2a}
\end{figure}

One simple and minimal way of ensuring these conditions is to impose
\begin{equation}
B_- = 0 .
\label{asymptotic_bc}
\end{equation}
In fact, using Eq.~\eqref{cond_impor}, this condition fixes the
frequency uniquely:
\begin{equation}
\label{omega_general}
\omega_\star
=
-\,\frac{i\lambda_+}{\Sigma_0}
=
-i\,
\frac{\lambda_+ \sigma^2}{Q_I R} \,\equiv\,-i \,\Gamma_{\rm rel} \,,
\hskip0.5cm\,\Gamma_{\rm rel}\ge0
\end{equation}
where in the third equality we used
$        \Sigma_0={Q_I R}/{\sigma^2} $  (see Eq.~\eqref{Sigma1_origin}).
Since $Q_I>0$, the frequency lies entirely in the lower half of the complex
$\omega$-plane, so that the corresponding
solutions remain bounded in time.  Two features make our results qualitatively different
from standard black hole quasi-normal modes:
\begin{itemize}
\item  With our convention for the harmonic time dependence,
Eq.~\eqref{harmonic_solution_a}, the frequency $\omega_\star$
is associated with  a single, purely damped mode. The perturbation
therefore describes relaxation, avoiding instabilities in time leading to exponential growth. Since $\lambda_+$ depends linearly on the multipole number $\ell$, the rate of relaxation is faster
for large $\ell$. We have
\begin{eqnarray}
R\,\omega_\star&=&-i \frac{\ell\,\sigma^2}{Q_I}\,,\hskip1.7cm 
{\ell \ge 1},\, 
{\rm spin-1\, modes}\,, \nonumber
\\
R\,\omega_\star&=&-i \frac{(\ell-1)\,\sigma^2}{Q_I}\,,\hskip0.9cm {\ell \ge 2},\, {\rm spin-2\, modes} \,.
\nonumber
\end{eqnarray}
Besides multipole number,
 the rate of relaxation depends on the properties
of the internal geometry -- the internal charge $Q_I$ and
the regularizing parameter $\sigma$. See Fig.~\ref{fig:AA2a}. These quantities can be varied
so to tune the value of  $\Gamma_{\rm rel}$, whose value is therefore controlled  by  properties of the interior geometry.
 Hence, even if the exterior geometry is 
Schwarzschild, the dynamics of fluctuations is very different 
from standard black holes, and depends on the particular
vector-tensor theory we are considering. 
\item Because  the boundary conditions lead to purely algebraic constraints, the  system under investigation does {\it  not} admit overtones. The latter  are in fact related  to interference effects in the two-way system of modes propagating in black-hole spacetimes -- a condition that is not realized in our setup, given the unidirectionality of the transport. 
\end{itemize}

\bigskip

Collecting  all three conditions together, we conclude that the corresponding mode function is continuous across the surface. It  has the
following analytic form on both  sides of the object surface:
\begin{equation}
\label{mode_function}
\Phi(t,r)
=
A\,
\exp\left[
\frac{\lambda_+}{\Sigma_0}
\left(
\Sigma_0 y(r)-t-\chi(r)
\right)
\right] .
\end{equation}

It remains to be checked whether the selected branch is also well behaved at large
radius. In the exterior region we have
\begin{equation}
\label{res_dele}
\Delta_1(r)=r+Q_E R \quad \hskip0.3cm
{\rm and} \hskip0.3cm \quad
\Sigma_1(r)
=
\frac{
\sqrt{Q_E^2R^2+2rQ_E R+2Mr}
}{
1-2M/r
} .
\end{equation}
Expanding the mode function for $r/R\gg 1$, we find
\begin{equation}
\label{mode_large_r}
\Phi(t,r)
\simeq
A\,e^{-i\omega t}
\left(\frac{r}{R}\right)^{\lambda_+}
\exp\left[
-\frac{2 \sqrt{2}\,\lambda_+\sigma^2\sqrt{Q_E+M/R}}{Q_I}
\sqrt{\frac{r}{R}}
\right] .
\end{equation}
Since for us $\lambda_+>0$, $Q_I>0$,  and 
$Q_E+M/R>0$, 
 the decaying  exponential factor dominates
over the growing power-law prefactor. The mode therefore decays exponentially as
$r\to\infty$. This provides an a posteriori consistency check that our
boundary conditions select a stable and normalizable
relaxation mode.

\paragraph{In what sense are our boundary conditions special?}
It is useful to pause and clarify in what sense the boundary conditions adopted
above are special and physically interesting. In fact, they are not the only mathematically admissible boundary
conditions for the problem. Rather, they define a minimal  prescription
breaking the chiral symmetry of the system,
 which leads to a distinctive and physically 
elegant relaxation spectrum with peculiar properties.

\medskip

\begin{itemize}
\item
We first consider the asymptotic condition~\eqref{asymptotic_bc}. As discussed
above, the general spectral condition depends on the two amplitudes
$(B_+,B_-)$. Therefore, for generic boundary data, the position of the pole is
not determined only by the background geometry and by the multipole number, but
also by the relative weight assigned to the two asymptotic branches. It is
possible that some choices of this relative weight also lead to modes with
${\rm Im}\,\omega<0$. However, this would amount to tuning the fluctuation data, so 
 to obtain a stable pole. By contrast, the condition
$        B_-=0
$ removes the potentially dangerous branch directly 
with no extra tunings.
 The resulting frequency is then fixed intrinsically by the
background quantities and by $\ell$, independently of any arbitrary
normalization of the perturbation. In this sense, Eq.~\eqref{asymptotic_bc} is the
minimal amplitude-independent prescription which selects the stable relaxation
branch.

\item
We then consider the condition imposed at the origin. As anticipated above,  the symmetry breaking prescription in the interior
geometry is essential to obtain a quantized spectrum. Our Neumann condition is a simple choice that breaks the chiral symmetry and leads to a physically interesting frequency spectrum. 
More generally, Robin-type conditions 
\begin{equation}
\label{robin_origin_general}
\left.
\left(
\partial_r \Phi + \kappa_\ell \Phi
\right)
\right|_{r=0}
=0
\end{equation}
would shift the frequency $\omega_\star$  of the single pole, Eq.\eqref{omega_general} (possibly also to real values for a complex $\kappa_\ell$). The choice of Robin condition
\eqref{robin_origin_general} in general depends on the pattern
of symmetry breaking one wishes to adopt.  
Our simple choice $\kappa_\ell=0$ is therefore a minimal physical 
condition in the original radial coordinate. 
\end{itemize}
  It would be interesting in the future to study more general possibilities, which would depend
on the symmetry-breaking physics under consideration. We will return in Section~\ref{sec:open_system} to touch this subject from a different perspective. 

\subsection{Black-hole limit, interior decoupling, and recovery of the chiral symmetry}
\label{sec:bh-limit}

\begin{figure}[t]
\centering
                \includegraphics[width=0.8\linewidth]{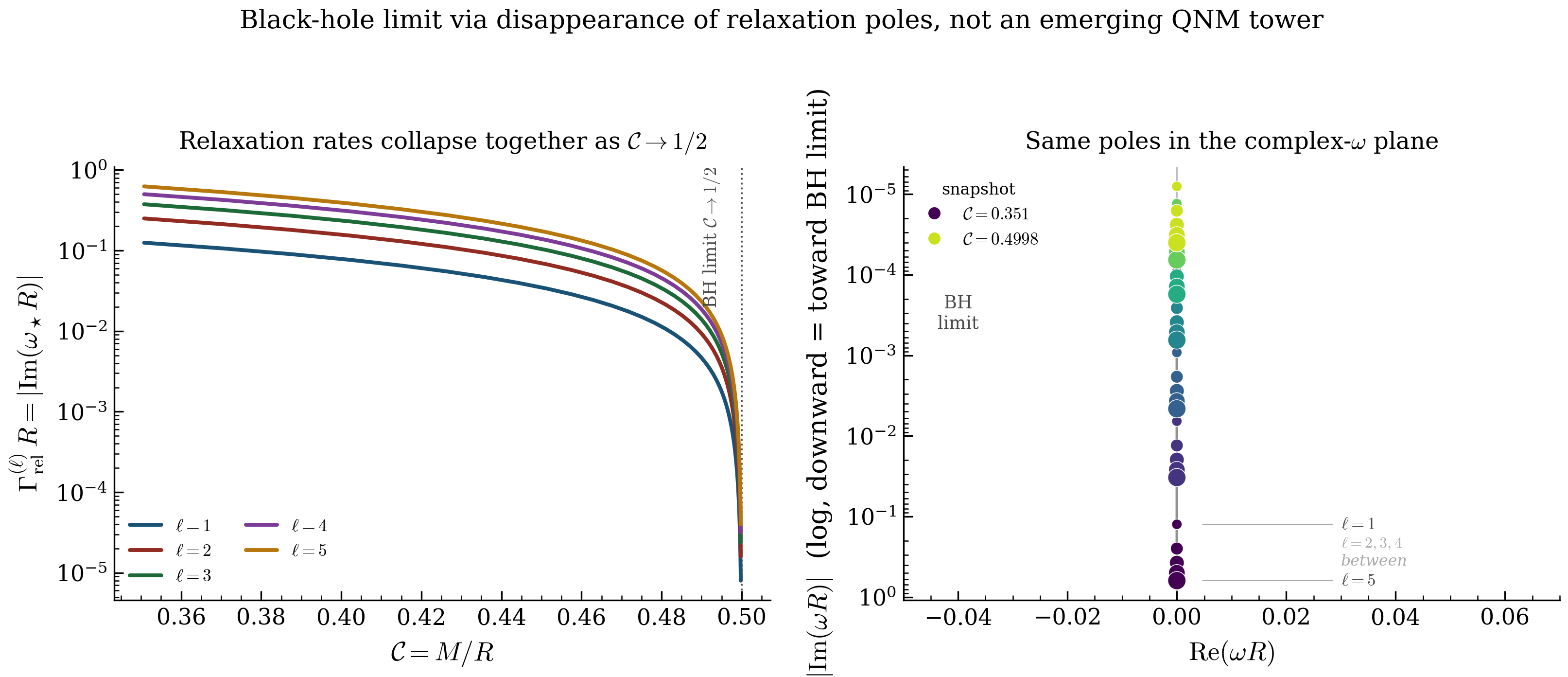}
\caption{\small \it  Trajectory of the relaxation poles $\omega_\star^{(\ell)}=-i\Gamma_{\rm rel}^{(\ell)}$ approaching the black-hole limit, along the family $\sigma\to0$, $\gamma=2+5/\sigma$ of Eq.~\eqref{bh_limit}. {\bf Left:} $\Gamma_{\rm rel}^{(\ell)}R$ versus compactness $\mathcal{C}$ for $\ell=1,\dots,5$, on a logarithmic scale: all multipoles collapse toward $\omega=0$ together as $\mathcal{C}\to1/2$ (dotted line), spanning several decades. {\bf Right:} the same poles in the complex-$\omega$ plane (logarithmic $|{\rm Im}\,\omega|$ scale) at several representative snapshots of increasing compactness (colour scale), showing the poles sinking toward the origin rather than spreading out into a finite-frequency tower. The black-hole limit is therefore approached through the \emph{disappearance} of the relaxation modes, not through the emergence of a conventional quasinormal spectrum.}
\label{fig:AA1bA}
\end{figure}
We now discuss the physical meaning of the black-hole limit.  %
It
corresponds to (see Section~\ref{sec_compt})
\begin{equation}
{\cal C}\to \frac{1}{2},
\qquad
\sigma\to0,
\qquad
\gamma\to\infty .
\label{eq:bh_limit_again}
\end{equation}
In this limit
$
\Sigma_0=\frac{Q_I R}{\sigma^2}
$
diverges, and therefore
\begin{equation}
\Gamma_{\rm rel}\to0,
\qquad
\tau_{\rm rel}
\equiv
\frac{1}{\Gamma_{\rm rel}}
\to\infty .
\label{eq:relaxation_time_diverges}
\end{equation}
Equivalently, all relaxation poles are pushed to the origin of the
complex-frequency plane,
$
\omega_{\star}\to0$. See Fig.~\ref{fig:AA1bA}. In this particular setup,
the black-hole limit is therefore not obtained by turning the relaxation modes
into an ordinary tower of Schwarzschild quasinormal modes. Rather, the
relaxation spectrum disappears from finite-frequency exterior dynamics.

This provides a useful way to interpret how the compact object approaches a
black hole. As \(\Sigma_0\) grows, the characteristic time needed for the
regular interior to communicate with the exterior grows without bound. On any
finite exterior time scale, the interior cannot complete its response. An
external observer therefore becomes insensitive to the object core, because its
own relaxation dynamics has been redshifted to arbitrarily late times. We refer
to this regime as the interior-decoupling limit.

The same conclusion can be phrased in terms of the chiral symmetry discussed in
Section~\ref{subsec:quadratic_action_correct}. 
The scale
$\Gamma_{\rm rel}$
introduced by the boundary conditions
is the physical remnant of the symmetry breaking induced by the
boundary-value problem.
%
In the black-hole limit this symmetry-breaking scale vanishes, and  the frequency range in which the interior response is dynamical
shrinks to the neighbourhood of \(\omega=0\).
  The
finite-frequency exterior theory therefore loses sensitivity to the mechanism
that had quantized the chiral data at finite compactness.
The explicit breaking
caused by the regular interior is pushed to zero frequency, and the
finite-frequency dynamics again resembles the unquantized chiral transport
behaviour.

\section{Green function, susceptibility, and tidal response}
\label{sec:green}

In the previous section we determined the intrinsic, source-free relaxation spectrum of the compact object. We now turn to its response to external perturbations. Because the odd-parity equations are exactly solvable, the retarded Green function can be obtained in closed form. This object contains more information than the pole spectrum alone: its residues determine the strength of the relaxation channels, its real-frequency imaginary part controls dissipation, and its zero-frequency limit encodes the static response.  (See e.g. \cite{DeLuca:2024ufn} and references therein.) In the present system the result is a Debye-type, purely relaxational susceptibility.

\subsection{Solutions of the homogeneous equations}
\label{sec_grsol}

Working in Fourier space, the retarded Green function is constructed from two homogeneous solutions, 
selected by physically motivated boundary conditions.
  Recall that  the solutions for the spin-1 and spin-2 modes, before imposing boundary conditions, have
the  general structure both in
 the exterior and the interior (see Eq.~\eqref{psi_int}):
\begin{align}
\psi(r)
&=
e^{-i\omega\chi(r)}
\left[
C_+ e^{\lambda_+ y(r)}
+
C_- e^{\lambda_- y(r)}
\right] ,
\label{psi_gen2}
\end{align}
We now proceed to impose the same type of boundary
conditions as in Section \ref{sec_bou}.
 
\paragraph{The regular solution.}
 We impose
the  Neumann boundary condition at the origin:
\begin{equation}
\partial_r \psi\big|_{r\to 0} = 0\,.
\label{eq:central_bc}
\end{equation}
 Normalising so that $C_+ = 1$ gives the regular solution
\begin{equation}
\psi_{\rm reg}(r;\omega)
=
e^{-i\omega\chi(r)}
\left[
e^{\lambda_+ y(r)}
+
q(\omega)\,e^{\lambda_- y(r)}
\right],
\label{eq:psi-reg}
\end{equation}
with coefficient
\begin{equation}
q(\omega)
=
-e^{(2 \ell+1)\, y_0}
\frac{\lambda_+ - i\omega\Sigma_0}{\lambda_- - i\omega\Sigma_0},
\qquad
y_0 \equiv  y(0)  ,
\qquad
\Sigma_0 \equiv \Sigma_1(0) .
\label{eq:q-def}
\end{equation}
The frequency dependence of $q(\omega)$ encodes
how the interior boundary condition selects a specific linear combination of
the two branches. The resulting $\omega$-dependence is  responsible
for the pole structure of the Green function, and will play a key
role in the following discussion. 

\paragraph{The passive solution.}
The second solution is selected by the asymptotic boundary condition at
spatial infinity.  As established in Section~\ref{sec:spectral}, the
characteristics of the system point inward, so spatial infinity is an inflow
boundary rather than an outflow boundary.  The physically convenient  condition for our
purposes 
is passivity: the object should respond to incoming perturbations without
spontaneously generating an exponentially growing mode in time.

As in Section  \ref{sec_bou}, the passive solution
we select is
\begin{equation}
\psi_{\rm pass}(r;\omega) = e^{-i\omega\chi(r)}\,e^{\lambda_+ y(r)} .
\label{eq:psi-pass}
\end{equation}
 As we will learn, the corresponding
Green function has poles only in the lower
half-plane, consistent with the conditions of causality and stability.

\subsection{The retarded Green function}
\label{subsec:gf_explicit}

We now construct analytically the retarded Green function using the two
homogeneous solutions \(\psi_{\rm reg}\) and \(\psi_{\rm pass}\) introduced
above.  In frequency space, the Green function is defined by the factorized
operator
\begin{equation}
\left(\mathcal J_\omega-\lambda_+\right)
\left(\mathcal J_\omega-\lambda_-\right)
G^R(r,r';\omega)
=
\delta(r-r') ,
\label{eq:GF_def_corrected}
\end{equation}
where we have suppressed the spin label \(j\).  Since the coefficient of
\(\partial_r^2\) in Eq.~\eqref{eq:GF_def_corrected} is \(\Delta_1^2\),
integrating across \(r=r'\) gives
\begin{equation}
\Delta_1^2(r')
\left[\partial_r G^R\right]_{r=r'_-}^{r=r'_+}
=
1 .
\label{eq:Green_jump_derivative_corrected}
\end{equation}
Equivalently, because \(G^R\) is continuous at \(r=r'\), this can be written as
the flux condition
\begin{equation}
\Delta_1(r')
\left[
\mathcal J_\omega G^R
\right]_{r=r'_-}^{r=r'_+}
=
1 .
\label{eq:Green_jump_flux_corrected}
\end{equation}

The Green function is therefore
\begin{equation}
G^R(r,r';\omega)
=
\frac{
\theta(r'-r)\,
\psi_{\rm reg}(r;\omega)\psi_{\rm pass}(r';\omega)
+
\theta(r-r')\,
\psi_{\rm pass}(r;\omega)\psi_{\rm reg}(r';\omega)
}
{\Delta_1(r')\,\mathcal W(r';\omega)} ,
\label{eq:GF_flux_W_corrected}
\end{equation}
where
\begin{equation}
\mathcal W(r';\omega)
=
\psi_{\rm reg}(r';\omega)\,
\mathcal J_\omega\psi_{\rm pass}(r';\omega)
-
\psi_{\rm pass}(r';\omega)\,
\mathcal J_\omega\psi_{\rm reg}(r';\omega)
\label{eq:Wronskian_corrected}
\end{equation}
is the corresponding first-order Wronskian. 
Substituting the explicit solutions of Section~\ref{sec_grsol}, one obtains
\begin{align}
G^R_j(r,r';\omega)
&=
\frac{
 e^{-i\omega[\chi(r)-\chi(r')]}
 e^{\lambda_+ y(r)-\lambda_- y(r')}
}
{\Delta_1(r')\,\left(\lambda_- - \lambda_+\right)}
\nonumber\\
&\quad\times
\left[
 e^{-(2\ell+1)y_0}
 \frac{\lambda_- - i\omega\Sigma_0}
 {\lambda_+ - i\omega\Sigma_0}
-
 \theta(r'-r)e^{-(2\ell+1)y(r)}
-
 \theta(r-r')e^{-(2\ell+1)y(r')}
\right] .
\label{eq:GF_explicit_correct}
\end{align}
This expression has a single pole at
\begin{equation}
\omega_\star=-i\Gamma_{\rm rel},
\qquad
\Gamma_{\rm rel}=\frac{\lambda_+}{\Sigma_0},
\label{eq:GF_pole_location_corrected}
\end{equation}
in agreement with the intrinsic relaxation frequency found in
Eq.~\eqref{omega_general}.
It is useful to isolate the pole-containing part of
Eq.~\eqref{eq:GF_explicit_correct}.  We define
\begin{equation}
\delta_\ell\equiv \lambda_+-\lambda_-=2\ell+1,
\qquad
\Delta\chi\equiv \chi(r)-\chi(r') .
\label{eq:delta_Deltachi_defs_corrected}
\end{equation}
Using
\begin{equation}
\frac{\lambda_- - i\omega\Sigma_0}
{\lambda_+ - i\omega\Sigma_0}
=
1+
\frac{\lambda_- - \lambda_+}
{\lambda_+ - i\omega\Sigma_0},
\label{eq:ratio_identity_corrected}
\end{equation}
the Green function can be decomposed exactly as
\begin{equation}
G^R(r,r';\omega)
=
G_{\rm an}^R(r,r';\omega)
+
G_{\rm pole}^R(r,r';\omega),
\label{eq:GF_exact_split_corrected}
\end{equation}
where the analytic transport contribution is
\begin{align}
G_{\rm an}^R(r,r';\omega)
 =
\frac{
 e^{-i\omega\Delta\chi}
 e^{\lambda_+ y(r)-\lambda_- y(r')}
}
{\Delta_1(r')\,\delta_\ell}
\,\left[
 \theta(r'-r)e^{-\delta_\ell y(r)}
+
 \theta(r-r')e^{-\delta_\ell y(r')}
-
 e^{-\delta_\ell y_0}
\right],
\label{eq:GF_an_part_corrected}
\end{align}
and the pole-containing contribution is
\begin{equation}
G_{\rm pole}^R(r,r';\omega)
=
\frac{
 e^{-i\omega\Delta\chi}
 e^{\lambda_+ y(r)-\lambda_- y(r')}
 e^{-\delta_\ell y_0}
}
{\Sigma_0\,\Delta_1(r')\,\left(\Gamma_{\rm rel}-i\omega\right)} .
\label{eq:GF_pole_exact_corrected}
\end{equation}
The split in Eq.~\eqref{eq:GF_exact_split_corrected} is exact.  Notice in
particular the last term in Eq.~\eqref{eq:GF_an_part_corrected}, proportional to
\(-e^{-\delta_\ell y_0}\).  This term is analytic in \(\omega\) and is required
for the exact decomposition of Eq.~\eqref{eq:GF_explicit_correct}.

Equivalently, in a neighbourhood of the pole one can write the Laurent form
\begin{equation}
G^R(r,r';\omega)
=
G_{\rm reg}^R(r,r';\omega)
+
\frac{\mathcal R(r,r')}{\Gamma_{\rm rel}-i\omega},
\label{eq:GF_Laurent_corrected}
\end{equation}
where \(G_{\rm reg}^R\) is analytic at
\(\omega=-i\Gamma_{\rm rel}\).  The residue is
\begin{align}
\mathcal R(r,r')
&=
\lim_{\omega\to -i\Gamma_{\rm rel}}
\left(\Gamma_{\rm rel}-i\omega\right)G^R(r,r';\omega)
\nonumber\\
&=
\frac{1}{\Sigma_0\,\Delta_1(r')}
\exp\left[
 -\Gamma_{\rm rel}\Delta\chi
 +\lambda_+ y(r)
 -\lambda_- y(r')
 -\delta_\ell y_0
\right] .
\label{eq:residue_correct}
\end{align}
The exact analytic remainder associated with the Laurent split is
\begin{align}
G_{\rm reg}^R(r,r';\omega)
&=
G_{\rm an}^R(r,r';\omega)
\nonumber\\
&\quad+
\frac{
 e^{\lambda_+ y(r)-\lambda_- y(r')}
 e^{-\delta_\ell y_0}
}
{\Sigma_0\,\Delta_1(r')}
\frac{
 e^{-i\omega\Delta\chi}
 -
 e^{-\Gamma_{\rm rel}\Delta\chi}
}
{\Gamma_{\rm rel}-i\omega} .
\label{eq:GF_reg_Laurent_corrected}
\end{align}
The second term in Eq.~\eqref{eq:GF_reg_Laurent_corrected} is regular at the
pole, since its numerator vanishes at
\(\omega=-i\Gamma_{\rm rel}\).  Thus the pole part is genuinely isolated, while
all remaining contributions are analytic in the neighbourhood of the relaxation
pole.

Fourier transforming the isolated Laurent pole in
Eq.~\eqref{eq:GF_Laurent_corrected} gives
\begin{equation}
G^R_{\rm pole}(t;r,r')
\supset
\theta(t)\,\mathcal R(r,r')\,e^{-\Gamma_{\rm rel}t} .
\label{eq:GF_time_Laurent_corrected}
\end{equation}
This is the characteristic signature of a relaxation channel: a perturbation
excited by a source decays as a pure exponential, with no oscillatory component.
If instead one Fourier transforms the exact pole-containing contribution
\eqref{eq:GF_pole_exact_corrected}, keeping the phase
\(e^{-i\omega\Delta\chi}\), the response is shifted along the characteristic,
\begin{equation}
G^R_{\rm pole}(t;r,r')
=
\frac{
 e^{\lambda_+ y(r)-\lambda_- y(r')}
 e^{-\delta_\ell y_0}
}
{\Sigma_0\,\Delta_1(r')}
\theta\!\left(t+\Delta\chi\right)
\exp\!\left[-\Gamma_{\rm rel}\left(t+\Delta\chi\right)\right] .
\label{eq:GF_time_exact_pole_corrected}
\end{equation}
The two descriptions are equivalent: the Laurent form isolates the local pole
residue, while the exact pole contribution also keeps the characteristic time
delay between the source and observation points.  In both cases the decay rate
is the same,
\begin{equation}
\tau_{\rm rel}^{-1}=\Gamma_{\rm rel} .
\end{equation}

Finally, note that \(y_0<0\) with the convention of Eq.~\eqref{def_y}.
Therefore the pole residue contains a factor \(\exp[-(2\ell+1)y_0]\), which
grows exponentially at large \(\ell\).  The coordinate-space multipole sum of
the pole contribution is consequently not controlled by the residue alone, but
by the full time-dependent combination together with the angular harmonic
kernel.  At sufficiently late retarded times the linear growth
\(\Gamma_{\rm rel}\sim \ell/\Sigma_0\) suppresses high multipoles
exponentially.  At very early times, instead, the pole contribution should be
understood as part of the full retarded Green function, including the analytic
transport term.


\subsection{Susceptibility and physical response}
\label{subsec:susceptibility}

The retarded Green function $G_j^R(r,r';\omega)$ 
contains the complete linear response of the
compact object.  However, in order to connect it to physically
measurable quantities, we must specify both how the system is
\emph{excited} and what is \emph{observed}.  Neither of these
corresponds to a sharp radial position. Rather, any realistic external
source distributes its coupling over a spatial profile, and any
realistic detector projects the resulting perturbation onto a smooth
test function.  The quantity that captures both aspects simultaneously
is the retarded susceptibility. We are going to show that in our context the
susceptibility acquires a  simple Debye-type relaxation profile.


\smallskip

The viewpoint we adopt is similar in spirit to the worldline effective-field-theory
approach to compact objects
\cite{Goldberger:2004jt,Goldberger:2005cd,Porto:2016pyg}, aimed
at calculating quantities such  as static and dynamical Love numbers.
One introduces a point-particle effective action in which the coupling
of the object to the ambient tidal field is encoded in a set of
frequency-dependent response kernels.  The poles of these kernels correspond to the
dynamical modes of the object, their imaginary parts at real frequency
govern energy absorption, and their zero-frequency limit determines the
conservative tidal deformability (see e.g.
\cite{Hui:2020xxx,Charalambous:2021mea}, as
well as the recent reviews \cite{Rodriguez:2026iot,Chakraborty:2026qru}) ~\footnote{
In the EFT framework   the response functions ultimately connect to observables. Our goal here  is to determine the response function itself, not its EFT matching.
}.

\paragraph{Definition of the linear observable.}

In order to define a sensible projection, we recall that the quadratic
action derived in Section~\ref{subsec:quadratic_action_correct} takes
the form
\begin{equation}
S^{(2)}
=
\frac{1}{2}\int dt\,dr\,\mu (r)
\left[(\mathcal{J}_t \Phi)^2+\Lambda \Phi ^2\right],
\end{equation}
with the weight $\mu (r)$ given explicitly by
\eqref{eq:correct_measure}, and $\Lambda$ in Eqs \eqref{def_olam}.  $\mu$ is the natural  measure of the 
space on which the unidirectional transport operator $\mathcal{J}_t$ acts.  
 We therefore introduce a smeared linear observable by
\begin{equation}
\mathcal{O}_f(t)
=
\int dr\,\mu (r)\,f(r)\,\Phi (t,r),
\label{eq:Of_def}
\end{equation}
where $f(r)$ is an arbitrary smooth probe (or detector) profile. Physically, $\mathcal{O}_f$ represents the signal measured by a
detector whose sensitivity kernel is $f(r)$,
weighted by the natural inner product measure $\mu(r)$.  Different
choices of $f$ probe different aspects of the perturbation. A profile
concentrated near the surface $r=R$ probes the near-surface dynamics,
while a profile peaked at large $r$ probes the asymptotic gravitational
field.  The freedom in $f$ is therefore the freedom in choosing what
observable we wish to measure. 

\paragraph{Linear response and the retarded susceptibility.}

Consider now adding to the action a small external source term that is time dependent, which 
couples linearly to the observable~\eqref{eq:Of_def},
\begin{equation}
\delta S_{\rm src}
=
\int dt\,J(t)\,\mathcal{O}_f(t)
=
\int dt\,dr\,\mu(r)\,J(t)\,f(r)\,\Phi(t,r).
\end{equation}
This represents, for example, an external tidal field whose spatial
profile is $f(r)$ and whose time dependence is $J(t)$.  The equation
of motion for $\Phi$ is now sourced by $ f(r) J(t)$.

In linear response theory, after Fourier transforming, the  value of $\mathcal{O}_f$
in the presence of the source $J$ is obtained
by the formula
\begin{equation}
\label{def_sus}
\mathcal{O}_f(\omega)
=
\chi_f^R(\omega)\,J(\omega),
\end{equation}
where $\chi_f^R(\omega)$ is called {\it retarded susceptibility}.
To compute it, note that the response of $\Phi$ to the source is
\begin{equation}
\Phi(t,r)
=
\int dt'\,dr'\,
G^R(r,r';t-t')\,f(r')\,J(t'),
\end{equation}
where $G^R(r,r';t-t')$ is the retarded Green function in the time
domain, which we discussed in Section \ref{sec:green}.  Substituting into the observable and Fourier-transforming,
we obtain Eq.~\eqref{def_sus}
 with
\begin{equation}
\chi_f^R(\omega)
=
\int dr\,dr'\,
\mu(r)\,
f(r)\,G^R(r,r';\omega)\,f(r').
\label{eq:chi_def}
\end{equation}
Equation~\eqref{eq:chi_def} is  a   projection of the
Green function over the probe profile. It is
analytic in the upper half of the complex frequency plane, as required
by causality~\footnote{
With our convention \(\mathcal L G^R=\delta(r-r')\), the measure appearing in the source term cancels against the measure multiplying the equation of motion. This is why only the observation integral carries the explicit factor \(\mu(r)\) in Eq.~\eqref{eq:chi_def}.}.

\paragraph{Pole structure and Debye form.} 
We now substitute the pole decomposition of the retarded Green
function:
\begin{equation}
G^R(r,r';\omega)
=
G_{\rm reg}^R(r,r';\omega)
+
\frac{\mathcal{R}_j(r,r')}{\Gamma_{\rm rel}- i \omega},
\qquad
\Gamma_{\rm rel}=\frac{\lambda_+}{\Sigma_0}>0,
\end{equation}
where $G_{\rm reg}^R$ is analytic near the pole, and
$\mathcal{R}(r,r')$ is the pole residue computed in
\eqref{eq:residue_correct}.
Inserting into \eqref{eq:chi_def} and separating the pole and
regular contributions, the susceptibility results
\begin{equation}
\chi_f^R(\omega)
=
\chi_{f,\rm reg}^R(\omega)
+
\frac{\mathcal{A}_f}{\Gamma_{\rm rel}- i \omega},
\label{eq:chi_pole}
\end{equation}
where the analytic part is
\begin{eqnarray}
\chi_{f,\rm reg}^R(\omega)
&=&
\int dr\,dr'\,
\mu(r)\,
 f(r)\,G_{\rm reg}^R(r,r';\omega)\,f(r'),
\end{eqnarray}
and the pole residue projected onto the profile $f$ is
\begin{eqnarray}
\mathcal{A}_f
&=&
\int dr\,dr'\,
\mu(r)\,
f(r)\,\mathcal{R}(r,r')\,f(r')
\\
&=&
\frac{e^{-(2 \ell+1)\,y_0}}{\Sigma_0}
\int dr\,dr'\,
\mu(r)\,
\frac{f(r)\,f(r')}{\Delta_1(r')}
\nonumber\\
&&\times
\exp\!\Big[
-\Gamma_{ \rm rel}\bigl(\chi(r)-\chi(r')\bigr)
+\lambda_+ y(r)
-\lambda_- y(r')
\Big].
%
\end{eqnarray}
In the second equality we use
 the explicit residue~\eqref{eq:residue_correct}.
  The factor $1/\Sigma_0$ originates from the pole residue of the
Green function; it controls the overall magnitude of the coupling
between the relaxation mode and the external source. The exponential factor $e^{-\Gamma_{\rm rel}(\chi(r)-\chi(r'))}$ encodes the
delay accumulated along the transport characteristic between source
point $r'$ and observation point $r$.
Notice that the double integral above is not symmetric in $r$ and $r'$, since the transport
problem has a preferred direction.

\bigskip

\begin{figure}[t]
\centering
                \includegraphics[width=0.6\linewidth]{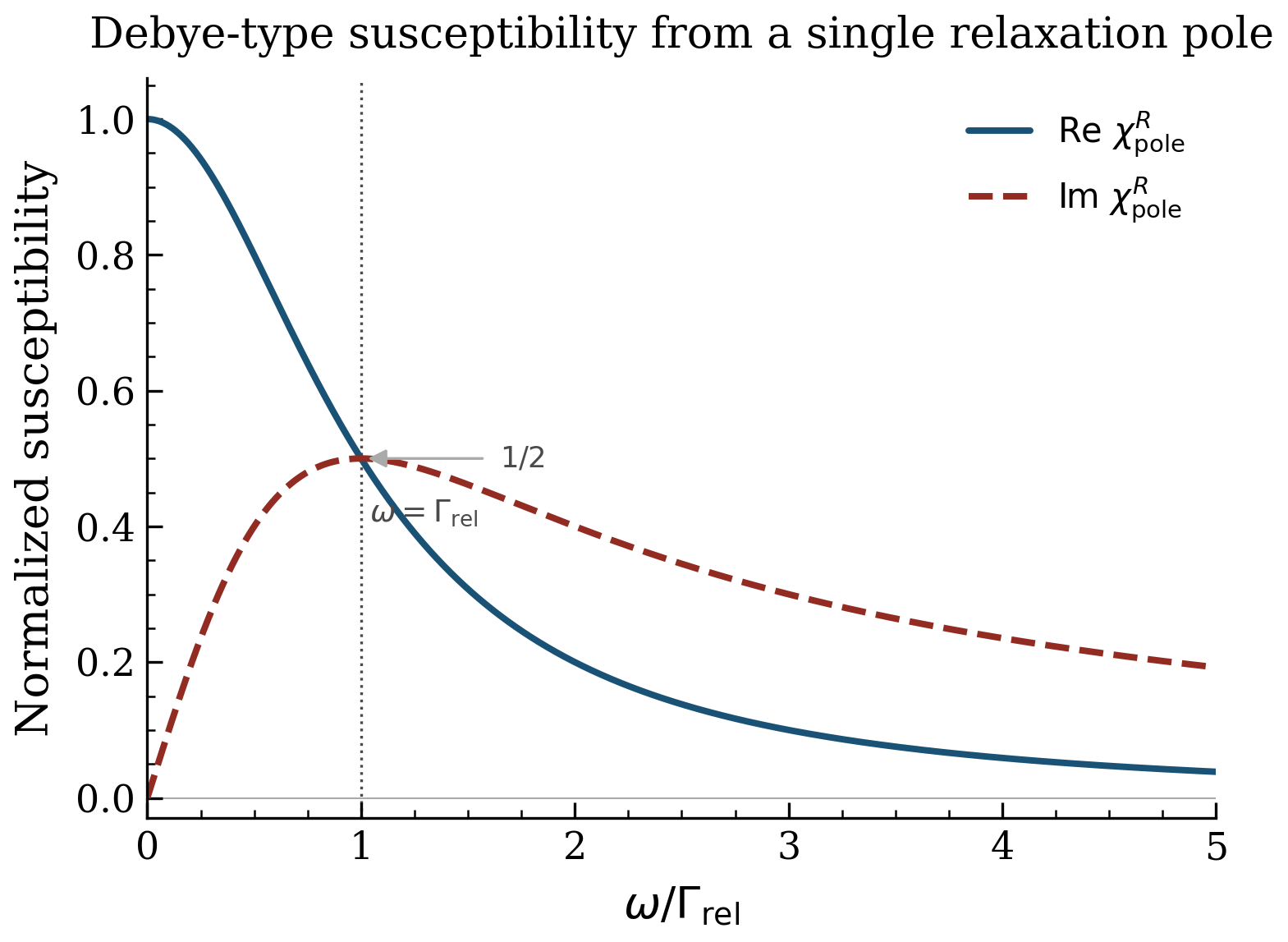}
\caption{\small \it 
Real  (solid) and imaginary (dashed)  parts of the pole contribution to the retarded susceptibility, Eq.~\eqref{eq:Debye_Re_Im}, plotted in the dimensionless normalization $\Gamma_{\rm rel}\chi^R_{f,\rm pole}/\mathcal A_f$ as functions of $\omega/\Gamma_{\rm rel}$.
  The response has the standard Debye form of a single overdamped degree of freedom: a Lorentzian real part of half-width $\Gamma_{\rm rel}$ centered at $\omega=0$, and an odd imaginary part controlling dissipation, crossing $1/2$ at $\omega=\Gamma_{\rm rel}$. No resonant peak at finite $\omega$ is present, consistent with the absence of an oscillatory quasinormal-mode tower in this sector. 
}
\label{fig:ABv1}
\end{figure}
The Debye form of the expression \eqref{eq:chi_pole} for the susceptibility  is the standard response function of a single
overdamped (non-oscillatory) degree of freedom. See
 Fig.~\ref{fig:ABv1}. 
 The
relaxation time is $\tau_{\rm rel}=\Gamma_{\rm rel}^{-1}$.  In our gravitational
setting, it is the timescale over
which the compact object returns to its unperturbed state after an external
perturbation, as already identified from the time-domain Green function
in Section~\ref{subsec:gf_explicit}.

Since $\mathcal{A}_f$ is real and
positive (for real and positive $f$), the real and imaginary parts of the pole contribution are
\begin{equation}
{\rm{Re}}\,\chi_{f,\rm pole}^R(\omega)
=
\frac{\mathcal{A}_f\,\Gamma_{\rm rel}}{\Gamma_{\rm rel}^2+\omega^2},
\qquad
{\rm{Im}}\,\chi_{f,\rm pole}^R(\omega)
=
\frac{\mathcal{A}_f\,\omega}{\Gamma_{\rm rel}^2+\omega^2}.
\label{eq:Debye_Re_Im}
\end{equation}
The real part is a Lorentzian peak centered at $\omega=0$, with half-width
$\Gamma_{\rm rel}$; it captures the conservative part of the response.  The
imaginary part, which is odd in $\omega$, governs dissipation.
 The fact that ${\rm{Im}}\, \chi_f^R(\omega)$ vanishes at $\omega=0$ is
consistent with the absence of an  absorptive response: a static
external field does not pump energy into the object.  
Energy absorption is a finite-frequency effect: the dissipative part turns on over the scale \(\Gamma_{\rm rel}\) and has no resonant peak at a non-zero oscillation frequency.
 This behavior is  different from a resonant
oscillator, whose imaginary part peaks near the oscillation
frequency rather than at zero (see e.g. \cite{kubo1991statistical}).

\subsection{Static response, and relation with Love-number interpretation}
\label{subsec:static_love_numbers}

Consider the conservative zero-frequency response for
the vector-type fluctuations.  In the vector
sector the two characteristic exponents $\lambda$ are given
in Eq~\eqref{exp_lam1}.
 The exterior static solution can therefore be written as
\begin{equation}
\psi_{\rm ext}(r)
=
B_+ e^{\lambda_+ y(r)}
+
B_- e^{\lambda_- y(r)} .
\label{eq:static_ext_branch_response}
\end{equation}
At large radius the two branches behave as \(r^\ell\) and \(r^{-\ell-1}\), since \(y(r)\sim\ln r\)
 (recall the considerations near Eq.~\eqref{res_dele}).   
It is therefore natural to regard $B_+$ as the applied tidal amplitude and
$B_-$ as the induced response amplitude.  We define the corresponding branch
response coefficient by
\begin{equation}
 \kappa^{(1)}_\ell
\equiv
{B_-\over B_+} .
\label{eq:kbranch_def}
\end{equation}
This quantity is closely related to a Love-number coefficient, up to the
normalization used to map the master variable to physical tidal and multipole
moments.  However, as emphasized below, it should {\it not} be confused with the
standard Schwarzschild black-hole Love number.

The ratio \eqref{eq:kbranch_def} follows directly from regularity at the
centre.  At $\omega=0$ the interior solution has the same characteristic form,
\begin{equation}
\psi_{\rm int}(r)
=
C_+ e^{\lambda_+ y(r)}
+
C_- e^{\lambda_- y(r)} .
\label{eq:static_int_branch_response}
\end{equation}
The Neumann condition at the centre gives
\begin{equation}
\left.\partial_r\psi_{\rm int}\right|_{r=0}=0
\quad\Longleftrightarrow\quad
\lambda_+ C_+ e^{\lambda_+ y_0}
+
\lambda_- C_- e^{\lambda_- y_0}=0,
\label{eq:neumann_static_branch_response}
\end{equation}
where $y_0\equiv y(0)$.  Hence
${C_-/ C_+}
=
- \left({\lambda_+/ \lambda_-}\right)\,
\exp\left[(\lambda_+-\lambda_-)y_0\right]
$.
Smooth matching at $r=R$, with the same normalization of the two branches on
both sides of the surface, gives $B_-/B_+=C_-/C_+$.  Therefore
\begin{equation}
\kappa^{(1)}_\ell
=
-\exp\left[(2\ell+1)y_0\right]
{\lambda_+\over \lambda_-}
=
\exp\left[(2\ell+1)y_0\right]
{\ell\over \ell+1} .
\label{eq:kbranch_vector_result}
\end{equation}
This is the static branch response of the regular compact object.  All the
interior dependence is encoded in the characteristic depth $y_0$.

\smallskip
Finally, notice that the black-hole limit requires care.  On the compact-object branch
considered here, the formal black-hole limit sends $y_0\to0$ and
$\Sigma_0\to\infty$.  Equation~\eqref{eq:kbranch_vector_result} then gives
\begin{equation}
\lim_{\rm BH}\, \kappa^{(1)}_\ell
=
{\ell\over \ell+1},
\label{eq:kbranch_bh_limit}
\end{equation}
not zero.  This is not a contradiction with the vanishing of the standard
Schwarzschild Love numbers.  The calculation above imposes regularity at the
centre of a horizonless object and then takes a limiting background.  The
black-hole Love-number calculation instead imposes regularity, or ingoing
behaviour, at a future horizon from the outset.  Therefore the two results
do not necessarily coincide -- and in fact they do {\it not} do so in our 
specific case.
 The quantity \(\kappa^{(1)}_\ell\) should therefore be interpreted as a
static branch-response coefficient of the regular compact object, rather than
as the ordinary black-hole Love number.  Its non-vanishing limiting value is a
 diagnostic of the infrared sensitivity of the static response to the
interior boundary condition.

\section{Integrating out the interior: a boundary effective theory}
\label{sec:open_system}


Having constructed the retarded Green function and the corresponding
susceptibility, we now reformulate the response problem in a complementary way,
 connected  to the membrane paradigm \cite{Damour:1978cg,Thorne:1986iy,Price:1986yy}. Besides
 being physically intuitive, it gives an interesting perspective for studying the
  phenomenon of chiral symmetry breaking in our setup.

In our discussion in the previous sections, we learned 
that the specification of the interior part of the geometry is essential for
obtaining a discrete, purely damped spectrum of intrinsic fluctuations, and a Debye-type
response to an external input. What if we coarse-grain the system and analytically integrate out the interior?  We intend to do so in
this Section, deriving an effective boundary action that captures
the same physics of the interior we eliminate, and provides a new perspective for the dynamics of fluctuations and the breaking of chiral symmetry.

Before starting our derivation,
it is useful to motivate our approach and  interpret our construction in 
a broader context.
 When discussing  black holes, the membrane paradigm replaces the true event
horizon by a fictitious timelike stretched horizon, endowed with effective
transport properties such as resistivity, conductivity and viscosity. This
description is not meant to introduce a material surface at the horizon.
Rather, it is an effective description for exterior observers: the inaccessible
region is integrated out and its effect is encoded in boundary conditions at a
timelike surface, placed infinitesimally outside the horizon.

A similar philosophy is often adopted for horizonless compact objects. Since
the microscopic structure of the interior may be unknown, one parametrizes the
surface by an effective response, for example through a reflectivity,
impedance, tidal response, or through a frequency-dependent boundary condition. Such a
description allows one to study ringdown and absorption without specifying all details of the interior.

\smallskip

The construction we are going to develop here follows the same logic
of the membrane approach, but in a more restrictive way. We do
not postulate a phenomenological membrane response at the surface. Instead, given that the
entire geometry is known and regular,  the response at the surface is
obtained by {\it analytically  integrating out} the interior, so as to obtain localized
surface action. Since we learned that the physics of the interior
is essential for breaking the chiral symmetry and for determining the poles
of the Green function, we can re-interpret the previous results
in terms of the properties of the boundary action. 

\subsection{Defining a Dirichlet-to-Neumann map}

Since the interior solution is known analytically, the region \(r<R\)
can be integrated out exactly.  The result is an effective exterior
problem in which the influence of the regular interior is encoded in a
frequency-dependent boundary kernel at \(r=R\). This procedure builds
a connection between our setup and the membrane
paradigm \cite{Damour:1978cg,Price:1986yy,Thorne:1986iy}.  Mathematically this
kernel is the Dirichlet-to-Neumann map of the interior problem \cite{Calderon:1980}. It
relates the value of the perturbation at the surface to its canonical
radial flux at the same position.
  In analogy with wave propagation and the membrane
paradigm, we shall refer to this kernel as the interior impedance. 
Because the interior perturbation does not respond instantaneously to a
surface disturbance, the resulting impedance is not a local constant but a
retarded, frequency-dependent kernel: in the time domain, the flux at
\(r=R\) depends on the past history of the boundary field, so that integrating
out the interior produces a genuinely non-Markovian boundary condition.

\medskip

The terminology ``Dirichlet-to-Neumann map'' (DtN) is taken from the
mathematical theory of boundary-value and inverse problems. In that context,
one prescribes the value of a field on the boundary and measures the
corresponding normal flux. The resulting operator encodes the response of
the bulk region hidden behind the boundary. In the present problem we use
the same idea in a radial setting. 
  The DtN  map provides a precise way of integrating out
the interior and replacing it by a surface response for the exterior
perturbations.
%
%
%
 Focussing on harmonic perturbations
$\Phi_j(t,r)=e^{-i\omega t}\psi_j(r)
$ the DtN map
 corresponds to  extracting the  quantity
\begin{equation}
\mathcal Z_{\rm int}(\omega)
=
\left.
\frac{\mathcal J_\omega[\psi_{\rm int}]}
{\psi_{\rm int}}
\right|_{R}
\label{eq:Zint_def}
\end{equation}
in Fourier space. The function $\mathcal Z_{\rm int}$ is the interior
impedance,  dictated by the boundary conditions for the mode functions. It weights the (continuous) flux at the object surface
in terms of the corresponding mode function.
Since the regular interior solution can be written as Eq.~\eqref{eq:psi-reg}, we find
\begin{eqnarray}
\mathcal Z_{\rm int}(\omega)
&=&
\frac{
\lambda_+ 
+
\lambda_- q(\omega)
}{
1+
q(\omega)
},
\label{eq:Zint_exact}
\\
&=&\frac{
\lambda_+\lambda_-(1-\rho_\ell)
+
i\omega\Sigma_0(-\lambda_+ + \rho_\ell\lambda_-)
}{
\lambda_- - \rho_\ell\lambda_+
-
i\omega\Sigma_0(1-\rho_\ell)
}.
\label{eq:Zint_rational}
\end{eqnarray}
where  we use the fact that  \(y(R)=0\), and
in the second line we
 introduced the combination
$ 
\rho_\ell
\equiv
e^{(2 \ell+1)\,y_0}
$.

\smallskip

The quantity $\mathcal Z_{\rm int}$ in the
 Robin condition \eqref{eq:Zint_def} leads to   an effective boundary action.
Recall the quadratic action \eqref{eq:quadratic_action_correct}  derived in Section \ref{subsec:quadratic_action_correct}.  The boundary variation of the
exterior problem, expressed in Fourier space,  is proportional to

\begin{equation}
\delta S_{\rm ext}\big|_{R}
=
-\int\frac{d\omega}{2\pi}\,
\mu(R)\Delta_1(R)\,
\mathcal J_\omega[\Phi_{\rm ext}(\omega)]_R\,
\delta\Phi_R(-\omega).
\end{equation}
where the sign follows from the fact that \(r=R\) is the inner boundary
of the exterior region. 
Therefore the effective boundary term in our setup may be written in frequency space
as
\begin{equation}
{
S_{\rm bdry}
=
\frac12
\int\frac{d\omega}{2\pi}\,
\mu(R)\Delta_1(R)\,
\Phi_R(-\omega)\,
\mathcal Z_{\rm int}(\omega)\,
\Phi_R(\omega),
}
\label{eq:Sbdry_sec6}
\end{equation}
where \(\Phi_R(\omega)\equiv\Phi_{\rm ext}(R,\omega)\).  
%
The
boundary term \eqref{eq:Sbdry_sec6} obtained by
integrating out the interior should be understood as a frequency-space, 
 linear response boundary condition -- not as an ordinary conservative action.
 (A fully causal variational
formulation would require the a Schwinger--Keldysh doubling in this
context, see e.g. \cite{Caron-Huot:2025tlq}.) 
It is useful to rewrite  it in coordinate space, in order to
make its non-local character explicit.  We define the time-domain kernel
\begin{equation}
{\cal Z}_{\rm int}(t-t')
=
\int\frac{d\omega}{2\pi}\,
e^{-i\omega(t-t')}\,
{\cal Z}_{\rm int}(\omega) .
\label{eq:Zint_time_kernel}
\end{equation}
Then Eq.~\eqref{eq:Sbdry_sec6} becomes
\begin{equation}
S_{\rm bdry}
=
\frac12\,
\mu(R)\Delta_1(R)
\int dt\,dt'\,
\Phi_R(t)\,
{\cal Z}_{\rm int}(t-t')\,
\Phi_R(t') .
\label{eq:Sbdry_time_nonlocal}
\end{equation}
Equivalently, the effective boundary condition at the surface may be
written as
\begin{equation}
\left.
{\cal J}_t\Phi_{\rm ext}(t,r)
\right|_{R}
=
\int dt'\,
{\cal Z}_{\rm int}(t-t')\,
\Phi_R(t') .
\label{eq:nonlocal_boundary_condition_time}
\end{equation}

This form makes manifest that integrating out the interior produces a
memory term at the surface, which
fully encapsulates the physics of the interior.  The exterior perturbation at time \(t\) is
not determined only by the instantaneous value of the boundary field
\(\Phi_R(t)\), but by its past history through the kernel
\({\cal Z}_{\rm int}(t-t')\).  If the impedance ${\cal Z}_{\rm int}(\omega)$  were independent of
frequency, then
\({\cal Z}_{\rm int}(t-t')\) would be proportional to
\(\delta(t-t')\), and the boundary condition would reduce to a local Robin
condition at \(r=R\).  The genuine frequency dependence of
\({\cal Z}_{\rm int}(\omega)\) is the ingredient that makes the
effective boundary dynamics non-local in time.

\subsection{Global spectrum from impedance matching}
\label{subsec:impedance_matching}

We now re-derive the  relaxation spectrum of Section \ref{sec:spectral} from an impedance
matching condition, providing a transparent
alternative interpretation of the compact object dynamics.

The exterior perturbation, after imposing our passivity condition
$B_- = 0$ of~\eqref{asymptotic_bc}, takes the form
$
\psi_{\rm pass}(r;\omega) = e^{-i\omega\chi(r)}\,e^{\lambda_+ y(r)}
$. The impedance presented by the passive exterior solution at the surface
$r=R$ is
\begin{equation}
\mathcal{Z}_{\rm ext}(\omega)
\equiv
\frac{\mathcal{J}_\omega[\psi_{\rm pass}]}{\psi_{\rm pass}}
\bigg|_{R}
=
\lambda_+,
\end{equation}
This exterior impedance is {independent of frequency}:
the passive exterior simply presents a fixed impedance $\lambda_+$ at
the surface, regardless of $\omega$.  This is a consequence of the
transport character of the exterior: because both branches carry the
same phase $e^{-i\omega\chi}$, the flux-to-field ratio of the decaying
branch is purely determined by its spatial exponent $\lambda_+$.

The global perturbation problem requires the field and flux to be
continuous across $r=R$.   
This is equivalent to requiring that the ratio of flux to field
is the same on both sides:
\begin{equation}
\mathcal{Z}_{\rm int}(\omega)
=
\mathcal{Z}_{\rm ext}(\omega).
\label{eq_imp}
\end{equation}
 This is the \emph{impedance matching condition}.  A global mode exists
only when the interior and exterior present the same impedance at
the matching surface.  (In circuit-theory terms: the interior element
and the exterior transmission line are in resonance when their
impedances are equal.) 
%
%
Substituting the various
quantities involved, we find that the condition of 
Eq.~\eqref{eq_imp} requires to find the frequency
such that $q(\omega)=0$, which is
 \begin{equation}
{
\omega_\star
=
-\frac{i\lambda_+}{\Sigma_0}
=
-\frac{i\lambda_+\sigma^2}{Q_I R}.
}
\label{eq:global_pole_sec6}
\end{equation}
This is the same result as obtained directly
from the spectral condition in Section~\ref{sec:spectral} and from the
pole of the retarded Green function in
Section~\ref{sec:green}.
 The impedance matching condition $q(\omega_\star) = 0$ has an intuitive
physical interpretation.  
 Recall that \(q(\omega)\) is the coefficient of the second radial branch in
the regular interior solution.  In the passive exterior prescription this
branch is absent.  The condition \(q(\omega_\star)=0\) therefore means that, at
the relaxation frequency, the regular interior solution reduces to the same
single branch selected in the exterior.
%
 \begin{equation}
\psi_{\rm int}(r;\omega_\star) = e^{-i\omega_\star\chi(r)}\,e^{\lambda_+y(r)},
\end{equation}
which is identical in form to the passive exterior solution.  The field
therefore passes continuously through the surface $r=R$ without any
mismatch or ``reflection''.  The entire configuration --- interior and
exterior --- is described by the same functional form, with no abrupt
change at the surface.   
 This is the gravitational analogue of impedance matching in optics or
electronics, where the transmission coefficient becomes unity and
reflection vanishes when the impedances of two media are equal.
 See Fig.~\ref{fig:ABcD1}, left panel.
 
\subsection{Boundary response, symmetry breaking, and black-hole limit}
\label{subsec:bhlimit_Zint}
\label{subsec:boundary_kernel_symmetry}

\begin{figure}[t]
\centering
\includegraphics[width=0.47\textwidth]{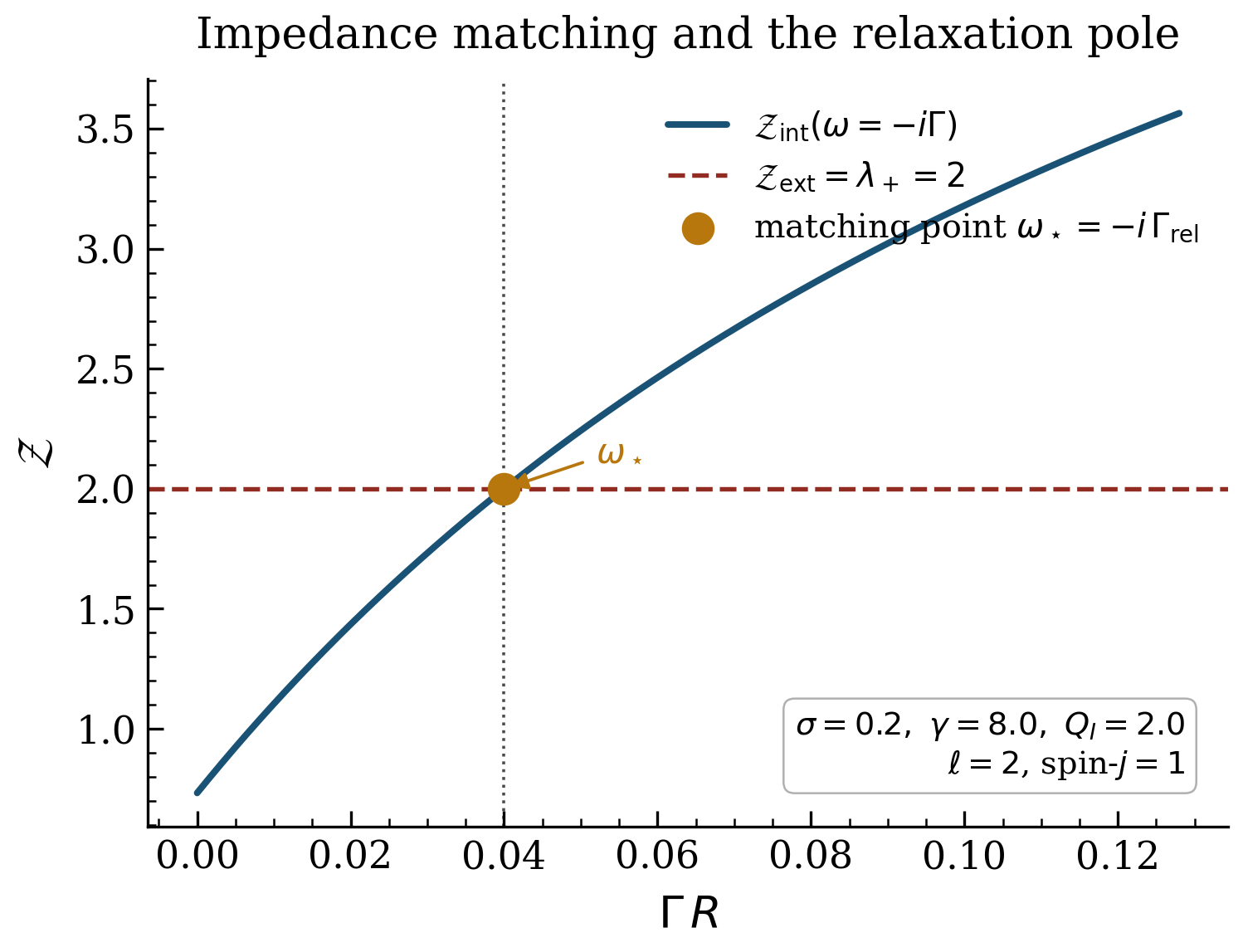}
\includegraphics[width=0.51\textwidth]{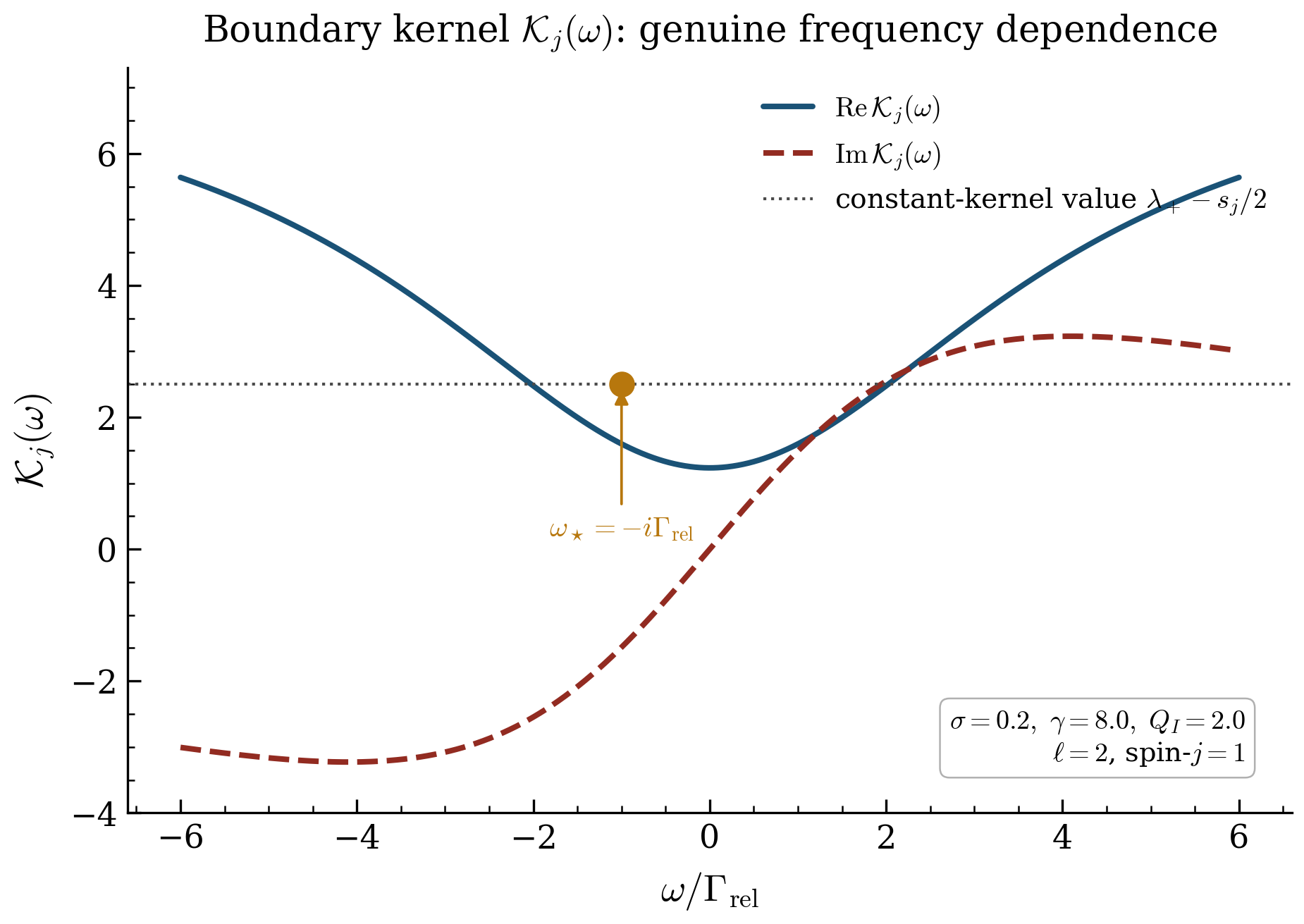}
\caption{\small \it {\bf Left:} 
 Impedance matching at the surface $r=R$ along the purely damped axis $\omega=-i\Gamma$. See Section \ref{subsec:impedance_matching}. The solid curve is the interior impedance $\mathcal{Z}_{\rm int}(\omega=-i\Gamma)$ obtained from the Dirichlet-to-Neumann map of the regular interior; the dashed line is the constant exterior impedance $\mathcal{Z}_{\rm ext}=\lambda_+$. The two curves cross at a single point (dot), the relaxation frequency $\omega_\star=-i\Gamma_{\rm rel}$, where the interior solution loses its subdominant radial branch and matches the passive exterior solution smoothly, with no reflection at $r=R$. 
  {\bf Right:} Real and imaginary parts of the effective boundary kernel
${\cal K}_j(\omega)={\cal Z}_{\rm int}^{(j)}(\omega)-s_j/2$, evaluated on the
real-frequency axis and plotted as functions of $\omega/\Gamma_{\rm rel}$.
The non-trivial frequency dependence of \({\cal K}_j\) shows that integrating
out the regular interior produces a time-nonlocal boundary condition for the
exterior transport problem. This real-frequency plot should be distinguished
from the pole condition: the relaxation mode is obtained by analytic
continuation to the imaginary-frequency axis, where impedance matching gives
\({\cal Z}_{\rm int}^{(j)}(\omega_\star)=\lambda_+\), equivalently
\(q(\omega_\star)=0\), and hence
\(\omega_\star=-i\Gamma_{\rm rel}\).
 In both plots we select  $\sigma=0.20$, $\gamma=8$, $Q_I=2$, $\ell=2$, spin sector $j=1$. }
\label{fig:boundary_kernel}
\label{fig:ABcD1}
\end{figure}

We close this section by connecting the  boundary response for
the coarse-grained system -- after integrating out the interior --
with the framework of chiral 
symmetry breaking as developed in the previous sections.  
As we discussed, the bulk
factorized equations have an enlarged chiral symmetry along the characteristic
coordinate $\nu$.  This symmetry is a property of the local transport problem.
The compact object, however, is defined by a global boundary-value problem:
regularity at the centre, matching at the surface, and the retarded prescription
in the exterior.  Once the interior is integrated out, all this information is
encoded in the frequency-dependent surface response.

\smallskip

To
summarize what we learned so far in this Section: the exact Dirichlet-to-Neumann map of the interior can be written as
$\left.
{\cal J}_{\omega}\psi_{{\rm int}}
\right|_{R}
=
{\cal Z}^{(j)}_{\rm int}(\omega)\,
\psi_{{\rm int}}(R)$, 
where
$
{\cal J}_{\omega}
=
\Delta_1\partial_r+i\omega\Sigma_1 
$. 
The function ${\cal Z}^{(j)}_{\rm int}(\omega)$ 
 is fixed by solving the regular
interior problem and evaluating the corresponding flux at the surface,
for each spin index $j=1,2$.
It is useful to express the same statements in terms of the field $\Psi_j$
introduced in Section~\ref{subsec:quadratic_action_correct}, defined by
\begin{equation}
\Phi_j =e^{s_j  y/2}\Psi_j .
\end{equation}
Since the surface is at $y=0$, the boundary condition becomes
\begin{equation}
\left.
\partial_y\Psi_j
\right|_{R}
=
{\cal K}_j(\omega)\,\Psi_j(R),
\label{eq:DtN_Psi_symmetry_section}
\end{equation}
with
\begin{equation}
{\cal K}_j(\omega)
=
{\cal Z}_{\rm int}^{(j)}(\omega)-\frac{s_j}{2}.
\label{eq:K_from_Z}
\end{equation}
Thus integrating out the interior replaces the smooth global problem by an
exterior transport problem supplemented by a non-local boundary condition at
$r=R$. See Fig.~\ref{fig:ABcD1}, right panel.

The symmetry-breaking pattern  is then interpreted as follows.  If
${\cal K}_j(\omega)$ were independent of frequency, Eq.~\eqref{eq:DtN_Psi_symmetry_section}
would be a local Robin condition in the adapted radial coordinate $y$.  Such a
condition fixes a radial branch or a fixed mixture of radial branches, but it
does not introduce a time scale along the chiral coordinate $\nu$.  It therefore
does not quantize the frequency.  By contrast, the actual kernel
${\cal K}_j(\omega)$ is frequency dependent.
This leads
 to a simple symmetry breaking pattern. The  frequency dependence preserves translations
along $\nu$, since the kernel depends only on the conjugate frequency.  However,
it breaks dilatations and special conformal transformations of $\nu$, because
these transformations rescale or mix frequencies.  At the level of the global
$SL(2,{\rm R})$ subgroup discussed in Section~\ref{subsec:quadratic_action_correct},
the boundary response therefore reduces the symmetry as
\begin{equation}
SL(2,{\rm R})
\quad
\longrightarrow
\quad
{\rm translations\ in}\ \nu .
\label{eq:breaking_pattern_symmetry_section}
\end{equation}
In this sense the relaxation pole is the dynamical remnant of the boundary
breaking of the chiral bulk symmetry.

\section{Outlook}
\label{sec:discussion}

In this work we have presented an analytically tractable compact object whose
odd-parity response is qualitatively different from the usual ringdown picture.
Although the exterior metric is exactly Schwarzschild, the non-trivial vector
profile supporting the regular interior changes the global perturbation problem.
The spin-1 sector and the homogeneous spin-2 sector are governed by factorized
first-order operators, and their local dynamics is one-way transport along a
single characteristic direction.  As a consequence, the Schwarzschild exterior
alone does not define a conventional quasinormal-mode problem in these sectors.
Once the regular interior, the matching conditions, and the asymptotic
prescription are included, a chiral symmetry of the parity-odd equations is broken
and the response is quantized.  For the minimal stable boundary prescription
studied here, each multipole carries a single purely dissipative pole, 
 with no oscillatory component and no overtone tower.  The same pole appears in
the exact retarded Green function, in the Debye-type susceptibility, and in the
frequency-dependent boundary kernel obtained after integrating out the regular
interior.

A useful lesson is that a Schwarzschild exterior does not uniquely determine the
dynamical response of a compact object.  The response is sensitive to how the
exterior region is glued to the interior, even when the background geometry
outside the surface is indistinguishable from that of a black hole.  In the
present model this information is encoded in a
  boundary kernel.
This kernel is not a phenomenological membrane parameter: it
is fixed by solving the regular interior problem and evaluating the flux at the
surface.  More generally, one may reverse the logic and regard such a kernel as
the coarse-grained observable data of the object.  Its pole structure, residues,
static response, and frequency dependence provide diagnostics of how the
interior stores, dissipates, and returns perturbations to the exterior.

This viewpoint also clarifies the role of symmetry.  The bulk odd-parity
equations possess a chiral symmetry acting on a characteristic coordinate, with a global \(SL(2,{\rm R})\) subgroup.  This symmetry is exact for
the local transport problem, but it is not a symmetry of the full global
response problem.  The boundary conditions, or equivalently the effective boundary kernel, reduce the symmetry to translations only,  and introduce
the relaxation scale \(\Gamma_{\rm rel}\).  It would be interesting to classify
more general boundary deformations of the chiral transport theory,
with different symmetry-breaking patterns.

A related direction is to make the symmetry breaking dynamical.  In the reduced
exterior description used here, the interior state is fixed and the boundary
kernel is a prescribed function.  This is analogous to a unitary-gauge
description of the gluing between the exterior and the regular interior.  A more
complete boundary effective theory could restore chiral covariance by promoting
the gluing datum to an edge degree of freedom.  After this edge variable is
integrated out, one should recover the non-local kernel derived in this work.
Such a formulation could clarify the relation between hidden symmetries,
boundary conditions, and the effective degrees of freedom associated with a
coarse-grained interior.

The black-hole limit provides another important direction.  Along the branch
considered here, in the black-hole
limit the relaxation poles collapse toward the origin of
the complex-frequency plane, and the finite-frequency exterior response
decouples from the interior.  The object approaches black-hole behaviour not by
developing the usual Schwarzschild quasinormal-mode tower, but by pushing its
own relaxation dynamics to arbitrarily long time scales.  At the same time, the
strict static response can remain sensitive to the regular-centre boundary
condition.  The apparent non-commutativity between the black-hole limit and the
zero-frequency limit deserves further study, since it may distinguish
dynamical decoupling at finite frequency from full equivalence to a
Schwarzschild black hole.

\smallskip

The present analysis has been classical and deterministic.  A natural extension
is an open-system formulation in which exterior perturbations are treated as the
system and the regular interior as an environment.  In a Schwinger--Keldysh
description, integrating out the interior should generate a causal influence
functional localized at the surface.  Its dissipative part would be related to
the Dirichlet-to-Neumann map computed here, while its noise kernel would depend
on the quantum or statistical state of the interior.  Such a framework would
allow one to study fluctuation--dissipation relations, decoherence, entropy
production, and possible coarse-grained response entropies associated with
families of interiors that lead to the same exterior boundary data.

\smallskip

Several technical extensions are worth pursuing.  First, the full coupled
spin-1/spin-2 odd-parity system should be analyzed, by
using the analytical Green functions provided, rather than only the
spin-1 sector and the homogeneous spin-2 equation.  Second, the even-parity
sector should be developed, since its constraints and mixing with the matter
fields may lead to a richer response.  Third, it would be useful to repeat the
construction for broader classes of vector--tensor compact objects, in order to
identify which features are model-dependent and which follow universally from
chiral transport and boundary-induced symmetry breaking.  Finally, the
susceptibility and Green function derived here should be connected to waveform
observables.  This would make it possible to ask whether relaxation without
ringdown can be constrained, or perhaps identified, in gravitational-wave data.

\smallskip

Overall, the main message is that compact objects with black-hole-like exteriors
can possess dynamical responses that are not small deformations of black-hole
ringdown.  In the example studied here, the regular interior breaks a hidden
chiral symmetry and replaces oscillatory quasinormal ringing by purely
dissipative relaxation.  Understanding this interplay between symmetry,
boundary response, and coarse-grained interior dynamics may provide a useful
route toward a broader theory of compact-object response beyond General
Relativity.

\subsection*{Acknowledgments}
 
 It is a pleasure to thank Katsuki Aoki, Naritaka Oshita, Shinji Mukohyama, and Shogo Tomizuka for useful discussions on related
 topics. GT is partially funded by the STFC grant ST/X000648/1,
 by the Royal Society grant 
IES\textbackslash{}R3\textbackslash{}243186 and  by the
 Leverhulme
Trust grant
RF-2026-166\textbackslash{}9.

  {\small

\providecommand{\href}[2]{#2}\begingroup\raggedright\endgroup

}

\end{document}